\documentclass[onecolumn,11pt]{IEEEtran}
\usepackage[lmargin=.8in,rmargin=.8in,tmargin=1in,bmargin=1in]{geometry}
\usepackage{colortbl}

\usepackage{amsmath}
\usepackage{amssymb}
\usepackage{amsfonts}
\usepackage{setspace}
\usepackage{cite}
\usepackage{ifthen}
\usepackage{epsfig}
\usepackage{array}
\usepackage{enumerate}
\usepackage{subfig}
\usepackage{tikz}
\usepackage{xspace}
\usepackage{hyperref}

\newtheorem{definition}{Definition}
\newtheorem{thm}{Theorem}

\newtheorem{lem}[thm]{Lemma}
\newtheorem{cor}[thm]{Corollary}
\newtheorem{remark}{Remark}
\newtheorem{example}{Example}

\newcommand{\beq}{\begin{equation}}
\newcommand{\eeq}{\end{equation}}
\newcommand{\bea}{\begin{eqnarray}}
\newcommand{\eea}{\end{eqnarray}}
\newcommand{\bean}{\begin{eqnarray*}}
\newcommand{\eean}{\end{eqnarray*}}
\newcommand{\bit}{\begin{itemize}}
\newcommand{\eit}{\end{itemize}}
\newcommand{\ben}{\begin{enumerate}}
\newcommand{\een}{\end{enumerate}}
\newcommand{\blem}{\begin{lem}}
\newcommand{\elem}{\end{lem}}
\newcommand{\bthm}{\begin{thm}}
\newcommand{\ethm}{\end{thm}}
\newcommand{\bpf}{\begin{proof}}
\newcommand{\epf}{\end{proof}}

\newcommand{\supth}{\textrm{th}}
\newcommand{\comment}[1]{}
\newcommand{\codeTableRowspace}{1.2}
\newcommand{\advMSR}[4]{In the absence of security requirements ($p=0$), repair of any node $f~(1\leq f \leq #1)$ is performed by connecting to any $d=#2$ nodes and downloading the inner product of the two symbols in each node with $[1~~~x_f]$, where $x_f$ is the $f^{\supth}$ element of $\mathbf{x}=[0~1 #3~2~6~5~4]$. The repair of any node $f~(1 \leq f \leq #1)$, with security from compromise of any $p=1$ node, is performed by connecting to any $(d+2p)=#4$ nodes and downloading the inner product of the two symbols in each node with $[1~~~x_f]$. The code is optimal: it requires the minimum possible storage (it is MDS), and for this storage, it requires the minimum possible amount of download in each of the aforementioned scenarios.}
\def\colortablecell{}

\def\setJ{{\text{\!\tiny$\mathcal{J}$}}}
\def\bigsetJ{{\text{$\mathcal{J}$}}}
\def\setI{{\text{\!\tiny$\mathcal{I}$}}}
\def\bigsetI{{\text{$\mathcal{I}$}}}
\def\matT{V}

\title{Information-theoretically Secure Erasure Codes \\ for Distributed Storage}
\author{Nihar~B.~Shah, K.~V.~Rashmi, Kannan~Ramchandran, {\em Fellow, IEEE}, and P.~Vijay~Kumar, {\em Fellow, IEEE}
\thanks{Nihar~B.~Shah, K.~V.~Rashmi and Kannan~Ramchandran are with the Dept. of EECS, University of California, Berkeley, CA 94703, USA. Email: \{nihar,\,rashmikv\,kannanr\}@eecs.berkeley.edu. P.~Vijay Kumar is with the Dept. of ECE, Indian Institute Of Science, Bangalore, India. Email: vijay@ece.iisc.ernet.in. P. Vijay Kumar is also an adjunct faculty member of the Electrical Engineering Systems Department at the University of Southern California, Los Angeles, CA 90089-2565. } \thanks{This paper was presented in part at Globecom 2011~\cite{ourGlobecom2011} and in part at ISIT 2012~\cite{ourISIT2012}.}
}

\begin{document}
\maketitle\thispagestyle{empty}
\begin{abstract}
%
Repair operations in distributed storage systems potentially expose the data to malicious acts of passive eavesdroppers or active adversaries, which can be detrimental to the security of the system. This paper presents erasure codes and repair algorithms that ensure security of the data in the presence of passive eavesdroppers and active adversaries, while maintaining high availability, reliability and efficiency in the system. Our codes are optimal in that they meet previously proposed lower bounds on the storage, network-bandwidth, and reliability requirements for a wide range of system parameters. Our results thus establish the capacity of such systems. Our codes for security from active adversaries provide an additional appealing feature of `on-demand security' where the desired level of security can be chosen separately for each instance of repair, and our algorithms remain optimal simultaneously for all possible levels. The paper also provides necessary and sufficient conditions governing  the transformation of any (non-secure) code into one providing on-demand security.
\end{abstract}

\section{Introduction}\label{sec:intro}
The amount of data stored in large-scale, distributed storage-systems such as data centers is increasing exponentially.
In order to scale massively at low costs, data-centers employ inexpensive commodity hardware. As these components are prone to failure~\cite{ford2010availability,rashmi2013hotstorage}, the system must possess enough redundancy to ensure that data remains reliable and available in the face of these failures. One means of introducing redundancy is via replication. However, replication is inefficient with respect to storage space utilization, and thus in order to scale economically, data centers are increasingly turning to the use of erasure coding as a far more efficient option~\cite{ghemawat2003google,hdfs_codes_blog}.

Consider a distributed storage system with $n$ storage nodes across which some data (termed the \textit{message}) is to be stored. Each of these $n$ nodes stores only a fraction of the data. In order to provide reliability and availability, the erasure codes considered in this paper ensure that a user (termed a \textit{data collector})  must be able to recover the message from the data stored in \textit{any} $k$ of the $n$ nodes. This property is called \textit{data reconstruction} or simply \textit{reconstruction}. The reconstruction property provides the storage system a capability of tolerating failures of \textit{any} $(n-k)$ of the $n$ nodes.

Upon failure of a storage node, a \textit{replacement} node is designated to store the data that was stored previously in the failed node. The replacement node recovers this data by downloading (a part of) the data stored in the remaining nodes. This is termed a \textit{node-repair} or simply a \textit{repair} operation. Traditional erasure codes typically handle repair by first downloading and reconstructing the entire message and then extracting the required data from it. Such an operation is quite wasteful of the network resources~\cite{rashmi2013hotstorage,rashmi2014hitchhiker,asterisxoring}, and several recent works~\cite{rashmi2014hitchhiker,ourAllerton,ourITW,ourProductMatrix,ourAllertonJournal,
ourMISERJournal, papailiopoulos2013repairTransactions, tamo2013zigzag, cadambe2011polynomial, wang2012long,  suh2011journal,sasidharan2013high,gaston2012quasi,lin2014novel, li2013framework,kurihara2013generalization,hou2013basic,gerami2013optimized,tian2013exact,shah2013minimizing, jiekak2012cross,asterisxoring} propose new erasure codes and repair algorithms addressing this issue.

Security is an important aspect of distributed storage systems, and this is underscored by the many incidents of data compromise in the recent past (e.g.,~\cite{citibank2011hack,linkedIn2012hack1,yahoo2012hack}). This paper considers the problem of designing codes and algorithms for distributed storage that ensure security in addition to maintaining other properties such as reliability, availability and efficiency. We consider the information-theoretic notion of security (and not the computational notion), making no assumption about the computational power of the adversary.

In designing a secure distributed storage system, special attention must be given to the repair operations since they can pose security hazards. For instance, in a system using traditional erasure codes, the repair of a failed node would require the download of the entire data at the replacement node. An eavesdropper tapping onto the replacement node can thus obtain the entire data. Alternatively, during repair of a failed node, an active adversary that has captured one or more of the remaining nodes may pass corrupt data during the repair operation. Such an action would corrupt the data stored in the replacement node, and these errors may then propagate across the entire system during subsequent repairs of other nodes. This makes repair operations a serious security threat, and motivates the goal of this paper. 

\begin{figure*}
\subfloat[]{
\includegraphics[width=.45\textwidth]{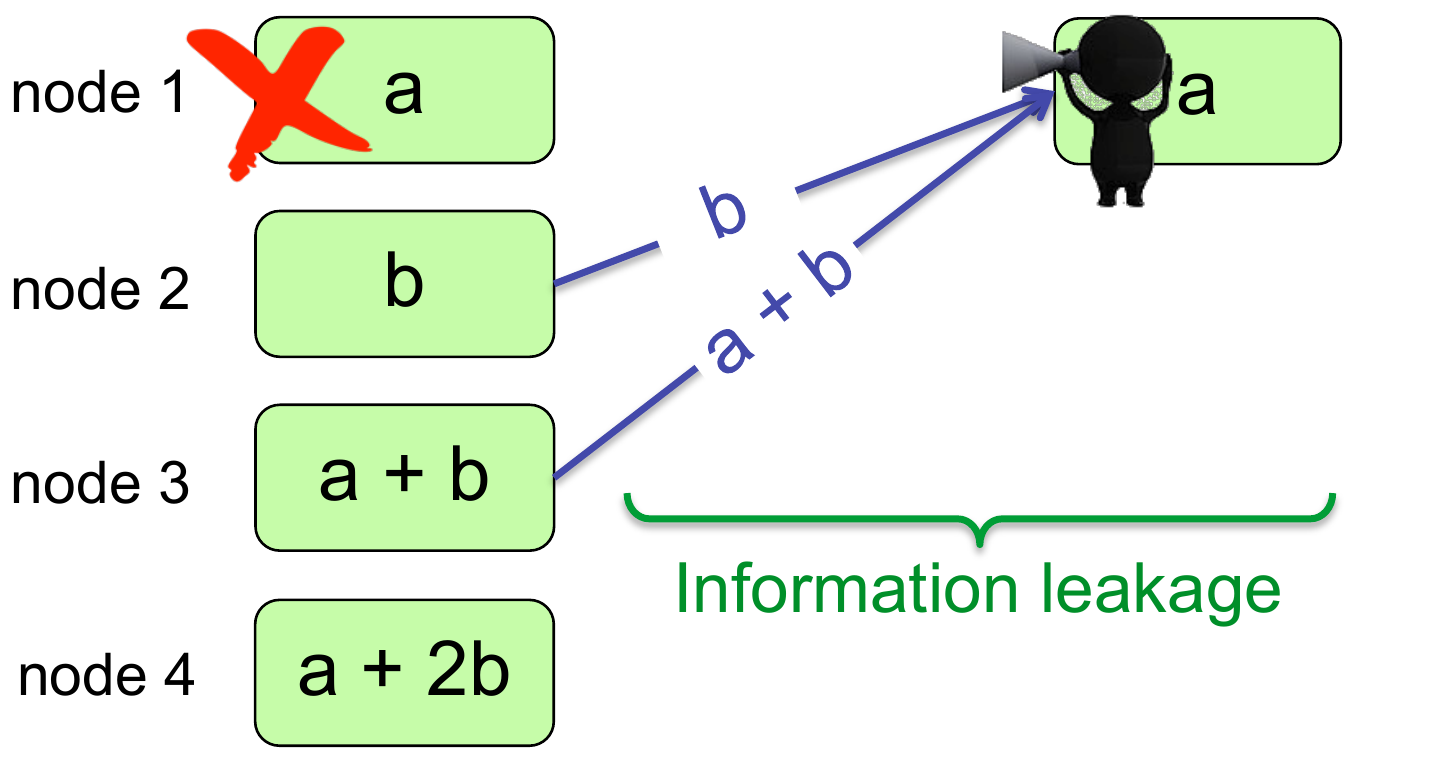}
\label{fig:eaves_RS}
}
\quad
\subfloat[]{
\includegraphics[width=.45\textwidth]{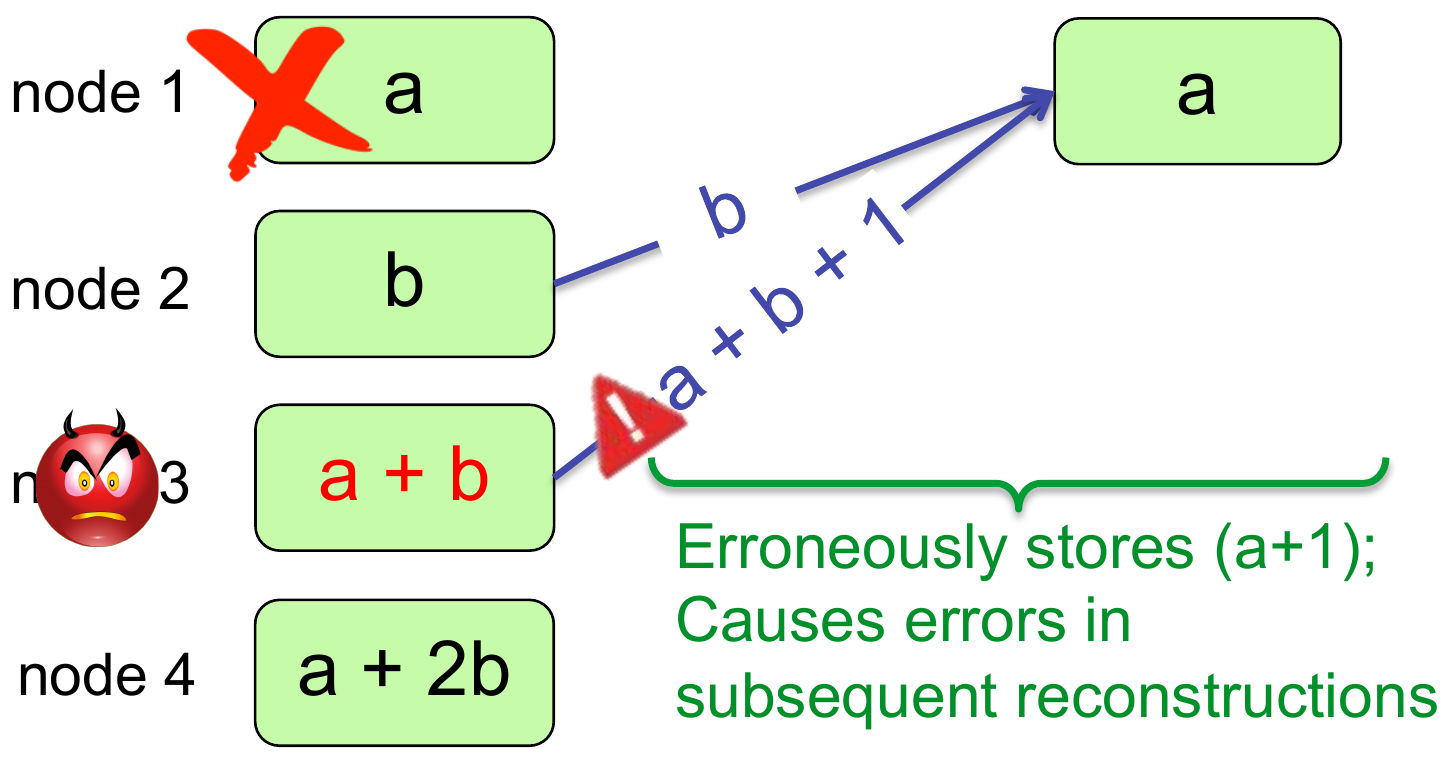}
\label{fig:adversary_RS}
}
\caption{Compromise of security when using a Reed-Solomon code due to (a) passive eavesdroppers, and (b) active adversaries.}
\label{fig:example_RS}
\end{figure*}

Fig.~\ref{fig:example_RS} illustrates the problem of security under repair dynamics using a toy example of a Reed-Solomon code with $n = 4$ and $k=2$. The code stores the message $\{a,b\}$ across four nodes, where both $a$ and $b$ belong to the finite field $\mathbb{F}_5$ in this example. It is easy to verify that the entire message can be recovered by downloading the data stored in any $k=2$ of the four nodes. On failure of any node, the replacement node connects to any two other nodes, downloads all data stored in them, from which the replacement node recovers the data stored in the node prior to failure. Fig.~\ref{fig:eaves_RS} illustrates a setting with passive droppers. Specifically, consider an eavesdropper who can read all the data stored in node $1$. Under the Reed-Solomon code, the eavesdropper can gain access to the symbol $a$. Furthermore, if the eavesdropper was also able to listen to the data passed for any repair operation, then it gains access to the entire message. Fig.~\ref{fig:adversary_RS} illustrates a setting with active adversaries. Suppose an active adversary gains access to one of the nodes, say node $3$, and suppose the repair of node $1$ is performed by connecting to nodes $2$ and $3$. Then, as seen in the figure, the adversary can pass malicious data in the repair process (for instance, $(a+b+1)$ instead of $(a+b)$), making the replacement node $(a+1)$ believing it is actually storing $a$. This error, in turn, propagates during further repairs, and also sabotages all subsequent reconstruction operations involving node $1$.

In this paper, we model the distributed storage system based on the \textit{regenerating codes} model introduced by Dimakis et al.~\cite{dimakis2010network}. In addition to the parameters $n$ and $k$ introduced earlier, this model has a third parameter $d$. Upon failure of a node, the replacement node may connect to \textit{any} $d$ of the remaining nodes, and should be able to recover the data that was stored previously in the failed node by downloading a minimal amount of data from these $d$ nodes. It was shown in~\cite{dimakis2010network} that there is a tradeoff between the total amount of data stored per node and that downloaded for repair of a failed node, and this is described in greater detail Section~\ref{sec:A_model}. The two extreme regimes in the tradeoff are the minimim storage regenerating (MSR) and the minimum bandwidth regenerating (MBR) regimes. The MSR regime aims to minimize the amount of storage required per node, and for this amount of storage, minimizes the amount of download. On the other hand, the MBR regime aims to keep the amount of download to an absolute minimum, and for this amount of download, minimizes the amount of storage. 

The initial work~\cite{dimakis2010network} on regenerating codes considers the repair to be `functional', wherein a replacement node need not be identical to the failed node, but should only satisfy further reconstructions and repairs. A strictly stronger and practical requirement is that of `exact' repair, where the replacement node must obtain and store data \textit{identical} to that in the failed node. Throughout this paper, we consider only exact repair.


The threat model we consider for security in this paper is an extension of the threat model proposed in~\cite{pawar2011securing} for the regenerating codes setup. Two classes of threats are considered:
\begin{itemize}
\item \textit{Security from passive eavesdroppers:} This threat class involves preventing leakage of any information about the data to passive eavesdroppers who may gain access to a subset of the storage nodes. By `passive' we mean that the eavesdropper may read and store any data it gains access to, but does not corrupt any data. In our threat model, two parameters $\ell$ and $m$ determine the level of security to be provided: the goal is to ensure that a passive eavesdropper having access to the data \textit{stored} in any $\ell$ of the nodes and additionally to the \textit{data passed for repair} in any $m$ of these $\ell$ nodes, gains zero information about the message.
\item \textit{Security from active (malicious) adversaries:} This threat class involves ensuring that the presence of an adversary who gains access to a subset of the storage nodes and wishes to actively corrupt the data does not affect the operations in the system. The active adversary may corrupt the data stored in a subset of storage nodes and also pass erroneous data to other nodes during their repair operations. In particular, for some parameter $p$ determining the level of security to be provided, the goal is to guard against $p$ nodes being corrupted by an active adversary. In such a scenario, if a replacement node (or a data-collector) connects to any of the $p$ corrupt nodes, then these corrupt nodes could pass arbitrary data in the repair (or data-reconstruction) operation. The goal here is to successfully accomplish the node-repair and data-reconstruction operations even in the presence of such an attack.
\end{itemize}

In this paper, we present explicit codes for distributed storage at MSR and MBR points that ensure security of the data from passive eavesdroppers and/or active adversaries. The secure codes at the MBR point are applicable for all values of the parameters $[n,\ k, \ d]$, and the secure codes at the MSR point are applicable for all values of the parameters satisfying $d \geq 2k-2$. 
The secure codes presented have two attractive features: 
\begin{itemize}

\item \textit{Optimality:} Our codes for security from active adversaries and from passive eavesdroppers at the MBR point are optimal for all values of the parameters. The codes for security from active adversaries at the MSR point is optimal for all parameters satisfying $d \geq (2k-2)$. 
The codes for security from passive eavesdroppers at the MSR point are optimal for all parameters satisfying $d \geq (2k-2)$ with $m \leq 1$. The codes presented in this paper, thus, establish the secure \textit{capacity} of such systems.  

\item \textit{On-demand security from active adversaries:} Under our codes, when dealing with active adversaries, the protection level $p$ need not be fixed a priori, but can be chosen flexibly at ``run-time'' when any repair or reconstruction operation is being executed. The protection level $p$ can be chosen separately and independently for each instance of repair or reconstruction. This ``on-demand'' security endows the system with the advantage of not having to preset the security level and associated system resources for the worst case. This is in contrast to the traditional models of information-theoretic security (e.g., \cite{pawar2011securing}) that take a ``static'' approach towards system design where the price is paid, in terms of the reduction in the size of the message stored, corresponding to the magnitude of the ``worst-case'' security level required. 
It turns out, perhaps surprisingly, that \textit{our codes do not not require any additional storage} to support this on-demand property, and are \textit{optimal for all values of $p$}.
\end{itemize}	


The problem at hand is closely related to the problem of secure network coding~\cite{lehman2004complexity,feldman2004capacity,jaggi2007resilient,koetter2008coding,silva2008rank,ngai2009secure,kosut2009nonlinear,yao2010network,cai2011secure}. The literature on secure network coding primarily considers a multicast setting where one or more receivers wish to obtain the entire data that was transmitted. The threat models typically consider the scenario of link compromise in the networks. The problem we consider, on the other hand, falls into the harder \textit{non-multicast} setting~\cite{lehman2004complexity}, and furthermore, needs to handle the harder case of adversaries or eavesdroppers being able to \textit{capture nodes} in the network~\cite{kosut2009nonlinear}. The results of this paper thus establish the capacity of a class of non-multicast networks in the presence of active adversaries or passive eavesdroppers having the ability to compromise nodes. Interestingly, the capacity-achieving codes that we propose in this paper are linear, deterministic and explicit.

While the primary focus of this paper is information-theoretic security, the results of this paper also are also relevant to other applications such as correcting network errors and erasures and for reducing latency of ``degraded reads'' in data centers. These applications are discussed in more detail in Section~\ref{sec:other_apps}.

The rest of the paper is organized as follows. Section~\ref{sec:model} formalizes the system model, and describes our approach towards the code construction problem. Some additional applications of our codes are also discussed in this section. Section~\ref{sec:literature} describes related literature. The codes presented in this paper build upon our previously proposed \textit{product-matrix framework}~\cite{ourProductMatrix}, and this framework is overviewed in Section~\ref{sec:PM}.  Sections~\ref{sec:A_MBR} and Section~\ref{sec:E_MBR} present explicit constructions of MBR codes that provide security from active adversaries and from passive eavesdroppers respectively. Section~\ref{sec:A_MSR} and Section~\ref{sec:E_MSR} present explicit construction of MSR codes that provide security from active adversaries and from passive eavesdroppers respectively. Section~\ref{sec:necessary} provides necessary and sufficient conditions governing the transformation of any (non-secure) erasure code into one providing on-demand security. Section~\ref{sec:conclusion} presents a concluding discussion.

\section{System Model, Our Approach, and Summary of the Results}\label{sec:model}
We will first describe the system model in the absence of security requirements (the `regenerating codes' model), following which we describe the extension that incorporates the provision of security. 
Alongside, we also describe our approach towards constructing secure distributed storage codes, and a summary of the results of this paper.

\subsection{Regenerating Codes Model (in the absence of security requirements)}\label{sec:RC}
Under the regenerating codes model~\cite{dimakis2010network}, the system comprises of $n$ storage nodes, across which data comprising of $B$ symbols  is to be stored. This set of $B$ symbols is called the \textit{message}, and each of these symbols is assumed to belong to a finite field $\mathbb{F}_q$ of size $q$. Each of the $n$ storage nodes has a capacity of storing $\alpha$ symbols. The data is to be stored such that a user (termed \textit{data-collector}) can recover the entire message by downloading the data stored in \textit{any} $k$ of these $n$ nodes. This process is termed \textit{data-reconstruction}, or simply \textit{reconstruction}. It follows that any system satisfying the reconstruction property can tolerate the failure of \textit{any} $(n-k)$ storage nodes without losing any data.

Now let us consider the repair operation. When a storage node fails, it is replaced by another node, called the replacement node, that stores exactly the same data as the failed node.  The regenerating codes model contains two additional parameters, $d$ and $\beta$, that are associated to the repair of failed nodes. The replacement node is permitted to connect to \textit{any} $d~(\geq k)$ nodes out of the remaining $(n-1)$ nodes while downloading $\beta~(\leq \alpha)$ symbols from each node. This set of $d$ nodes are termed the \textit{helper nodes} for this instance of repair. From the set of $d\beta$ symbols thus obtained, the replacement node must recover the $\alpha$ symbols that were stored in the failed node. The total amount $d\beta$ of data downloaded for repair purposes is termed the \textit{repair bandwidth}.

It is shown in~\cite{dimakis2010network} that the parameters associated with a regenerating code must necessarily satisfy the bound
\beq
B \leq \sum_{i=0}^{k-1} \min\left(\alpha,(d-i)\beta\right)~. \label{eq:cut_set_bound}
\eeq
Since both storage and bandwidth come at a cost, it is naturally desirable to minimize both $\alpha$ as well as $d\beta$, and try to achieve the bound~\eqref{eq:cut_set_bound} with equality, i.e.,
\beq
B = \sum_{i=0}^{k-1} \min\left(\alpha,(d-i)\beta\right)~. \label{eq:cut_set_bound_equal}
\eeq
It can be deduced (see~\cite{dimakis2010network}) that achieving~\eqref{eq:cut_set_bound_equal}, for fixed values of $B$ and $[n,~k,~d]$, leads to a tradeoff between the storage space $\alpha$ and the repair-bandwidth $d\beta$.\footnote{In addition to achieving~\eqref{eq:cut_set_bound_equal}, the parameters must also satisfy the following minimality condition: a reduction in either $\alpha$ or $\beta$ must result in a violation of~\eqref{eq:cut_set_bound}.}

Two important regimes of this tradeoff are its extremities, termed the minimum storage regenerating~(MSR) and minimum bandwidth regenerating~(MBR) regimes, described below. 

\subsubsection*{Minimum Bandwidth Regenerating (MBR) regime}
The MBR regime entails the minimum possible repair-bandwidth: the amount of data downloaded by a replacement node is no more than the amount of data that was stored in the failed node. To arrive at the MBR regime, one must first minimize the bandwidth $d\beta$ in the tradeoff~\eqref{eq:cut_set_bound_equal}, and then obtain the minimum storage $\alpha$ for this value of repair-bandwidth. Clearly, a replacement node must download at least as much data as what was stored in the failed node. Moreover, it is shown in~\cite{ourInterior_pts} that for optimal performance, every storage node must utilize its full storage capacity of $\alpha$ symbols. Thus, the minimum bandwidth required is $d\beta = \alpha$.
To satisfy~\eqref{eq:cut_set_bound_equal} with this value of $d\beta$, the parameters associated to the MBR regime must satisfy
\beq \alpha = \frac{B}{k} + \frac{(k-1)\beta}{2}~,\quad d\beta=\alpha\label{eq:MBR_parameters_initial}~. \eeq

\subsubsection*{Minimum Storage Regenerating (MSR) regime}
The MSR regime allows for the minimum possible storage capacity: the requirement of being able to recover all $B$ message symbols from any $k$ nodes mandates the storage per node to be at least $\frac{B}{k}$, and the MSR regime requires attainment of this minimum, i.e., has $\alpha=\frac{B}{k}$. Substituting this value in~\eqref{eq:cut_set_bound_equal}, it follows that the parameters associated to the MSR regime are
\beq \alpha = \frac{B}{k}~,\quad d\beta=\alpha+(k-1)\beta~. \label{eq:MSR_parameters_initial} \eeq
Any code operating in the MSR regime is thus Maximum-Distance-Separable (MDS).

The two extremities of the tradeoff curve have been well-studied in the literature, and there exist several explicit codes~\cite{ourAllerton,ourITW,suh2011journal,ourProductMatrix,ourAllertonJournal,ourMISERJournal,wang2012long,tamo2013zigzag, papailiopoulos2013repairTransactions,cadambe2011polynomial} operating in these regimes and achieving~\eqref{eq:cut_set_bound_equal} for exact repair. On the other hand, it is shown in~\cite{ourAllertonJournal} that at essentially all other points on the tradeoff curve, there \textit{cannot} exist any codes that satisfy~\eqref{eq:cut_set_bound_equal} for exact repair. Tighter outer bounds have been proposed for these intermediate points on the tradeoff curve~\cite{tian2013rate, sasidharan2014improved,mohajer2015new,prakash2015storage}.

Having described the basic setup of regenerating codes (in the absence of security), we will now present a description of the setting where security from passive eavesdroppers and/or active adversaries is required. 

\subsection{Security from Passive Eavesdroppers}\label{sec:E_model}
\subsubsection{Threat Model and Upper Bounds}~\label{sec:passive_exist}
We consider a threat model that extends the model introduced in~\cite{pawar2011securing}. under the model we consider, a passive eavesdropper may gain access to the data stored in a subset of the storage nodes, and possibly also to the data downloaded during repair of some of these nodes. We assume that the eavesdropper may possess unbounded computational power and has complete knowledge of the system protocol. It is required that such an eavesdropper should not be able to obtain any knowledge (in the information-theoretic sense) about the message. 

An eavesdropper can gain read-access to the data stored in any arbitrary set of at most $\ell~(< k)$ storage nodes. In addition, the eavesdropper may also listen to the data downloaded during (any number of) repair operations of some arbitrary subset of $m~(\leq \ell)$ of these $\ell$ nodes. We call such an eavesdropper an \textit{$\{\ell,\ m\}$-eavesdropper}, and term the security of data from such a passive eavesdropper as $\{\ell,\,m\}$-security from the passive eavesdropper. Note that since the data passed to a node for repair also contains the data it will finally store, the setting described above can be equivalently stated in the following manner:
\begin{definition}[$\{\ell,\,m\}$-security from the passive eavesdropper]
A code provides $\{\ell,\,m\}$-security from the passive eavesdropper if a passive eavesdropper gaining access to only the data stored in any $(\ell-m)$ nodes, and the data passed for repair to any $m$ nodes, obtains zero information about the message.
\end{definition}

As an example of this model, consider a peer-to-peer storage system. The $m$ nodes described above may represent nodes that are in a network belonging to the eavesdropper, thereby allowing the eavesdropper to listen to all the data downloaded when these $m$ nodes undergo (possibly multiple) failures and repairs across time. On the other hand, the $(\ell - m)$ nodes may represent the nodes which may be exposed only momentarily, allowing the eavesdropper access to only the data stored.




We now describe an upper bound on the size of the message that can be stored for any code under this system model. In~\cite{pawar2011securing}, Pawar et al. consider the setting of $m = \ell$ and provide an upper bound on the number of message symbols $B^*$ that can be stored in the information-theoretically secure system. It can be shown easily that even under our extended setting of having $m \leq \ell$, the proof of this upper bound (Theorem 1 of~\cite{pawar2011securing}) works for any value of $m~(\leq \ell)$.  Using the bound from~\cite{pawar2011securing} for this setting, it follows that the parameters associated to any regenerating code providing $\{\ell,\ m\}$-security from passive eavesdroppers must necessarily satisfy
\beq B^* \leq \sum_{i=\ell}^{k-1} \min\left(\alpha,(d-i)\beta\right)~. \label{eq:cut_set_bound_secure} \eeq
Specializing to the MBR regime of $\alpha=d\beta$, this bound becomes
\beq B^* \leq \left(k d  - \frac{k(k-1)}{2} \right)\beta - \left(\ell d - \frac{\ell(\ell-1)}{2} \right)\beta~.\label{eq:mbr_secret_cutset_ineq_intro}\eeq


For the MSR regime, the bound in~\eqref{eq:cut_set_bound_secure} turns out to be loose. Subsequent to the conference publication~\cite{ourGlobecom2011} of our code constructions for security from passive eavesdroppers, in~\cite{rawat2014optimal}, the authors have provided tighter upper bounds for the MSR regime:
\beq  B ^* \leq (k-\ell)(\alpha-\beta)~.\label{eq:msr_rawat_bound}\eeq


\subsubsection{Summary of Our Results for Passive Eavesdroppers} This paper provides explicit constructions for:
\begin{itemize}
\item MBR codes for all parameters $[n, \ k, \ d]$, $\{\ell,\ m\}$ providing information-theoretic security from passive eavesdroppers. These codes are optimal for all values of the parameters.
\item MSR codes for all parameters $[n, \ k, \ d \geq 2k-2]$, $\{\ell,\ m\}$ providing information-theoretic security from passive eavesdroppers. These codes are optimal whenever $m \leq 1$.
\end{itemize}

The MBR code is optimal by virtue of meeting the outer bound~\eqref{eq:mbr_secret_cutset_ineq_intro} (derived in~\cite{pawar2011securing}) with equality. The MSR code is optimal for $m \leq 1$ by virtue of meeting~\eqref{eq:msr_rawat_bound} (derived in~\cite{rawat2014optimal}).

Our codes thus establish the secure capacity of distributed storage systems for the case of passive eavesdroppers for all parameters at the MBR regime and for all $[n, \ k, \ d \geq 2k-2]$ and $m\leq 1$ at the MSR regime.

We take the following approach for constructing the codes. To construct a secure code for a given $[n,k,d]$, we choose a product-matrix (PM) code with identical values of system parameters $[n,k,d]$. In the input to the PM code (without secrecy constraints), we replace a carefully chosen set of \beq R = B - B^* \label{eq:number_random} \eeq message symbols with $R$ random symbols, where $B$ is as obtained by setting equality in~\eqref{eq:MBR_parameters_initial} for the MBR case and in~\eqref{eq:MSR_parameters_initial} for the MSR case, and $B^*$ is the number of message that the code can store under the specified security requirement. 
Each of these random symbols are chosen uniformly and independently from $\mathbb{F}_q$, and are also independent of the message symbols. We then prove that this construction ensures that no $\{\ell, m\}$-eavesdropper can obtain any information about the message, thus ensuring $\{\ell,m\}$-security from passive eavesdroppers.

%
%
%
%

\subsubsection{Example}
\begin{figure*}[t]
\centering
\includegraphics[height=1.6in]{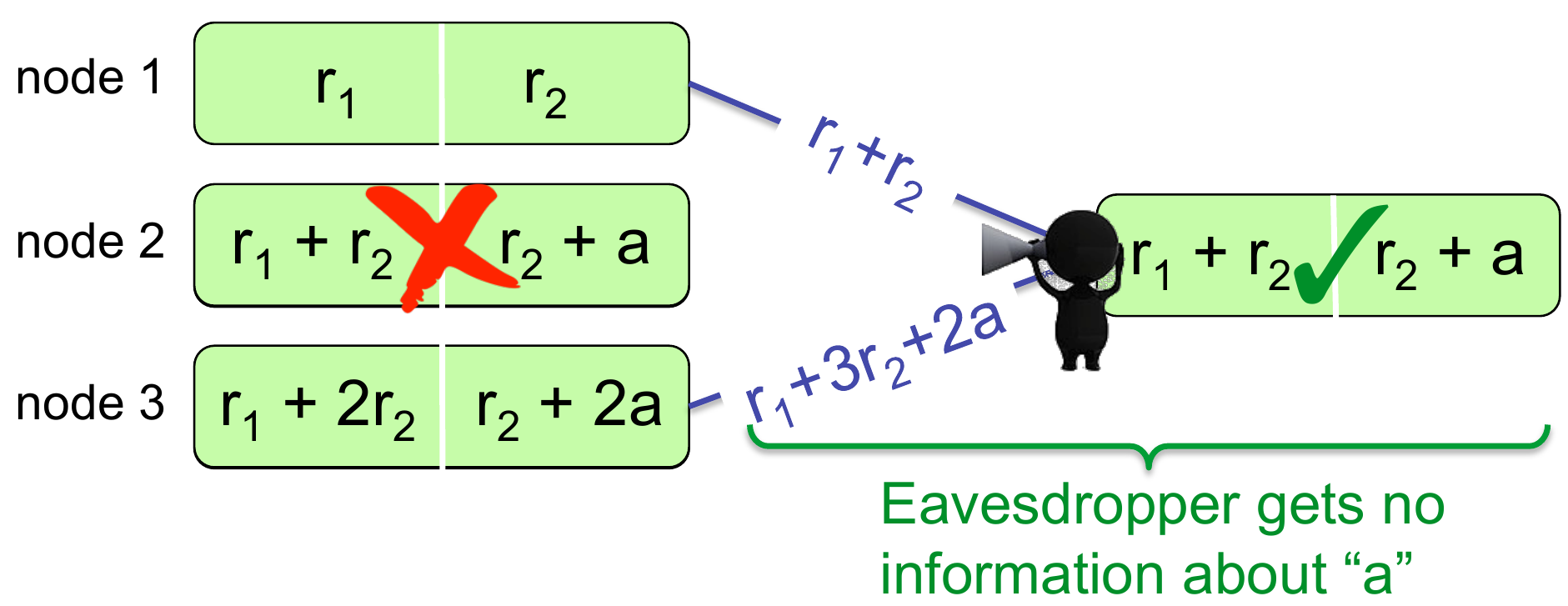}
\caption{MBR code with $[n=3,\ k=2,\ d=2]$ providing optimal $\{\ell=1,\ m=1\}$-security from passive eavesdroppers. Repair of node $2$ is shown, and one can see that the eavesdropper gains no information about $a$ from the data stored in or passed to any single node.}
\label{fig:example_eaves_MBR}
\end{figure*}

\newcommand{\nval}{3}
We illustrate our code construction with a toy example of an MBR code with $[n=\nval,\ k=2,\ d=2]$ providing $\{\ell=1,\ m=1\}$-security from passive eavesdroppers. The code is shown in Fig.~\ref{fig:example_eaves_MBR}. The message $\{a\}$ is encoded and stored across $n=\nval$ nodes in a manner that it can be recovered from any $k=2$ of the nodes. The alphabet of operation is the finite field $\mathbb{F}_{\nval}$. Symbols $r_1$ and $r_2$ are drawn uniformly at random from $\mathbb{F}_{\nval}$. A failed node is repaired by downloading one symbol each from any $d=2$ nodes. A passive eavesdropper gaining access to the data stored in any $\ell=1$ node and also to the data passed to that node during any of its repair operations ($m=1$) gets zero information about the message. For instance, the repair of node $2$ is shown in the figure, where one can see that the eavesdropper gains no information about $a$ from the data stored in or passed to any one node. The code is \textit{optimal} meaning that the amount of download required for any repair is the minimum possible, and furthermore, the amount of storage required in this setting is also minimum. 

\subsection{Security from Active Adversaries}\label{sec:A_model}
\subsubsection{Threat Model and Upper Bounds}
We consider the threat model wherein one or more nodes may be compromised to an active adversary. The adversary is assumed to be computationally unbounded and knows the protocol followed by the system. A node that is compromised to an active adversary may send arbitrarily corrupted data during any data-reconstruction or node-repair operation. It is required that the data-reconstruction operations complete successfully (without any errors) despite the presence of these active adversaries. This threat model was introduced in~\cite{pawar2011securing}. Note that this threat model does not require node repairs to be performed without errors, and only requires data-reconstruction operations to be performed without errors.

The upper bound on the size of the message that can be securely stored under the above described threat model with at most $p$ compromised nodes, as derived in~\cite{pawar2011securing} is
\beq
B \leq \sum_{i=2p}^{k-1} \min\left(\alpha,(d-i)\beta\right)~. \label{eq:cut_set_bound_secure_active}
\eeq 

Our codes achieve this bound and also ensure that every repair operation is also without corruption.
 
 Along the lines of traditional models of information-theoretic security, \cite{pawar2011securing} takes a ``static'' approach to secure-code construction, wherein the size of the message stored is reduced by an amount proportional to the magnitude ``$p$'' of the required security level. In particular, the value of $p$ chosen is typically based on worst-case estimation, which leads to considerable wastage of system resources during the normal operations. Furthermore, in the event that the number of nodes compromised is greater than that anticipated, the security level cannot be increased one-the-fly and hence security can no longer be guaranteed. 

Our approach to secure data from active adversaries is different from the conventional approach that we just described. Under our approach, the message is encoded and stored independent of any security requirements, and all the security requirements are handled ``on-demand'' at the decoding stage by downloading additional data. Let us assume that at some point in time we estimate a requirement of protection from the compromise of any $p$ out of the $n$ nodes in the network. Under our approach, this value of $p$ is \textit{not} fixed at the time of encoding of the data, but possible corruptions are corrected by downloading additional data during the reconstruction or repair processes. Now, during a particular instance of reconstruction or repair, the amount of additional data downloaded is a function of the desired protection level $p$, and this protection level can be chosen independently for each reconstruction and repair operation. This provides the advantage of not having to allocate resources for the worst case. This approach also enables the system to function with dynamic protection levels, and hence there is no need to estimate and fix the parameter $p$ beforehand: the system administrator is free to choose the desired level of security at runtime.

In our approach, the additional data downloaded for correcting possible corruptions is obtained by allowing a (greater) connectivity of $\Delta~(\geq d)$ nodes during repair and $\kappa~(\geq k)$ nodes during reconstruction. The parameters $\Delta$ and $\kappa$ depend on the desired protection level $p$ as
\begin{subequations} \begin{equation} \Delta=d+2p \end{equation} 
\begin{equation} \kappa = k+2p~.\end{equation} \label{eq:add_connect}\end{subequations}
Under this notation the outer bound of~\cite{pawar2011securing} translates to
\beq
B \leq \sum_{i=2p}^{\kappa-1} \min\left(\alpha,(\Delta-i)\beta\right)~. \label{eq:cut_set_bound_secure_active_translated}
\eeq
Observe that substituting the values of $\kappa$ and $\Delta$ in terms of $k$ and $d$ from~\eqref{eq:add_connect} in~\eqref{eq:cut_set_bound_secure_active_translated} results in~\eqref{eq:cut_set_bound}. It follows that achieving~\eqref{eq:cut_set_bound_secure_active_translated} with equality for all $p$ is equivalent to achieving~\eqref{eq:add_connect} and~\eqref{eq:cut_set_bound} with equality for all $p$. In particular, it amounts to achieving~\eqref{eq:add_connect} along with~\eqref{eq:MBR_parameters_initial} for the MBR regime and~\eqref{eq:MSR_parameters_initial} for the MSR regime.

Our goal is to construct codes which can secure the data from active adversaries for any value of the desired level of security $p$ that can be chosen separately at each instance of repair or reconstruction. We also require that, in addition, the codes be optimal simultaneously for all values of $p$. We term this \textit{on-demand security}.

\begin{definition}[On-demand security] A code is said to provide on-demand security if for \textit{any} arbitrary (feasible) choice of $p$ made at the time of a repair (or reconstruction) operation, the repair (or reconstruction) algorithm can detect and correct all corruptions in the presence of up to $p$ compromised nodes.
\end{definition}

Note that by all feasible values of $p$, we mean all values of $p$ for which the additional connectivity requirement described in~\eqref{eq:add_connect} can be satisfied (i.e., for all $p$ such that $(d+2p) \leq (n-1)$).

\begin{remark}[On-demand detection of errors]\label{rm:error_detect}
Note that the definition of on-demand security also requires \textit{detection} of corruption, when up to $p$ of the nodes are compromised during any repair (or reconstruction) operation. The corruption detection can be performed by connecting to $(d+p)$ (or $(k+p)$) nodes and downloading $\beta$ (or $\alpha$) symbols from each of them during any repair (or reconstruction) operation as opposed to $(d+2p)$ (or $(k+2p)$) nodes for correction of corruption. The error detection properties of the codes presented in this paper follow from their error correction properties in a straightforward manner using the basic relationship between detection and correction of errors in block codes~\cite{SloaneBook}. We hence focus only on the error correction in the rest of the paper, and the error detection properties follow as implicit corollaries.
\end{remark}


In our codes providing on-demand security, the parameters $[n,\,k,\,d]$ and $\{\beta,\,\alpha,\,B\}$ remain fixed, while the values of $\Delta$ and $\kappa$ may vary during different instances of reconstruction or repair operations depending on the security level $p$ desired at that time. Furthermore, our codes achieve the bound~\eqref{eq:cut_set_bound_secure_active_translated} with equality for all values of $p$, and are hence optimal.




\subsubsection{Summary of Our Results for Active Adversaries}
This paper presents explicit constructions for:
\begin{itemize}
\item MBR codes for all parameters $[n, \ k, \ d]$ providing on-demand security from active adversaries.  The codes are optimal for all values of the parameters.
\item MSR codes for all parameters $[n, \ k, \ d \geq 2k-2]$ providing on-demand security from active adversaries. The codes are optimal for all values of the parameters.
\end{itemize}
The optimality of the codes follows from the fact that the MBR codes meet~\eqref{eq:MBR_parameters_initial} and~\eqref{eq:add_connect} for all values of the parameters and the MSR codes meet~\eqref{eq:MSR_parameters_initial} and~\eqref{eq:add_connect} for all values of the parameters covered. Our results thus establish the secure capacity of distributed storage systems for the case of active adversaries for the aforementioned parameter regimes.

\subsubsection{Example}
\begin{figure*}[t]
\centering
\includegraphics[height=2.2in]{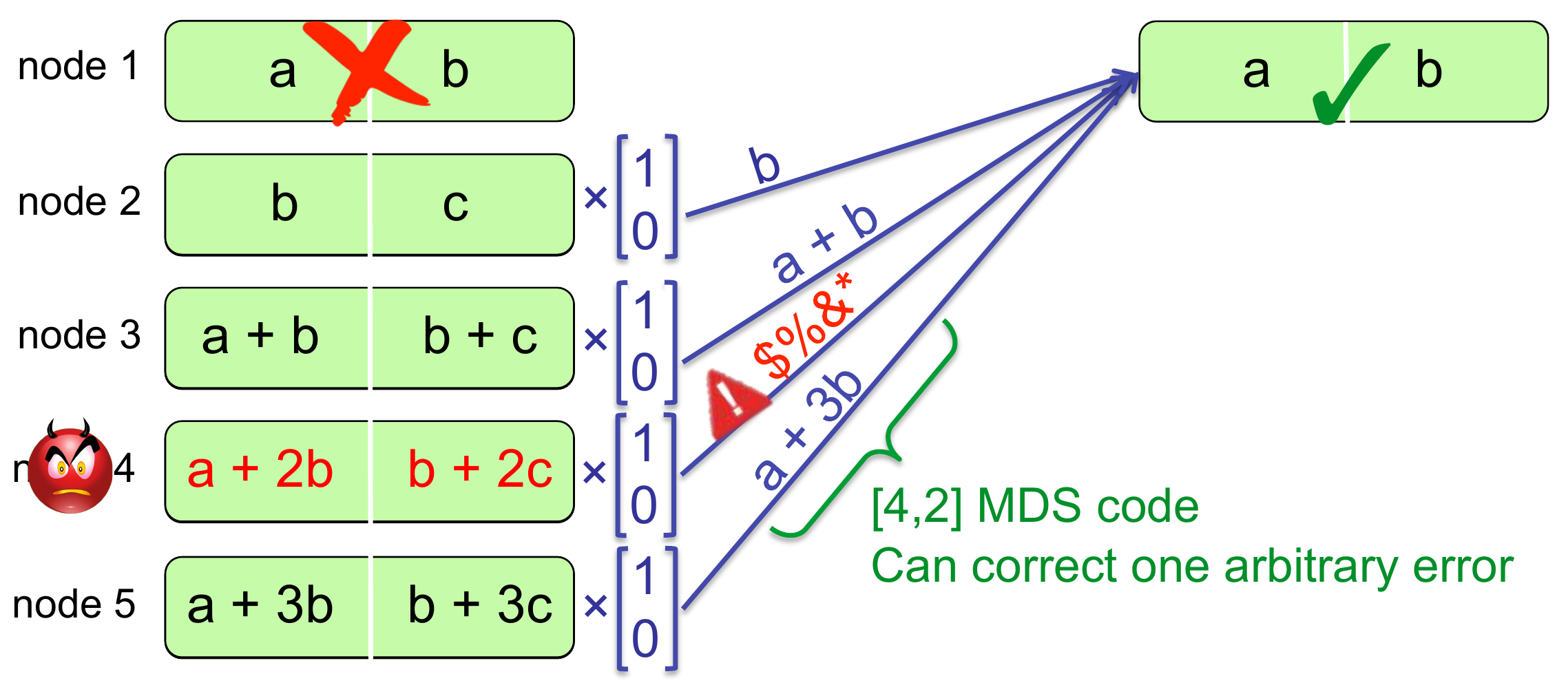}
\caption{MBR code with $[n=5,\ k=2,\ d=2]$ providing optimal and on-demand security. Repair of node $1$ is shown when security from at most $p=1$ compromised node is required. We can see that the repair operation proceeds successfully in spite of the presence of a compromised node that sends arbitrarily corrupted data. }
\label{fig:example_adversary_MBR}
\end{figure*}

We illustrate our code construction with an example of an MBR code with $[n=5,\ k=2,\ d=2]$ providing optimal and on-demand security. The code is shown in Fig.~\ref{fig:example_adversary_MBR}. The message $\{a,b,c\}$ is encoded and stored across $n=5$ nodes in a manner that it can be recovered from any $k=2$ of the nodes. The alphabet of operation is the finite field $\mathbb{F}_5$. In the scenario that a repair operation needs to be made secure from the compromise of any $p=1$ node, the replacement node connects to any $4$ nodes and downloads one symbol from each as shown in the figure. Here, even in the presence of one arbitrary corruption (node $4$ in the example), the desired symbols $a$ and $b$ are decoded correctly. When no security is required ($p=0$), a failed node is repaired by connecting to any $d=2$ nodes and downloading one symbol from each of them: the symbol passed by any node is identical to what it would have passed in the $p=1$ case. This code is optimal and provides on-demand security: a repair operation may choose the desired level of protection $p$ at runtime and the amount of network-bandwidth consumed in either case is the minimum possible, and furthermore, the amount of storage required to support this amount of download is also minimum.

\subsection{Other Applications}\label{sec:other_apps}
While the codes constructed in this paper are presented from the perspective of providing security in a distributed storage system, they can also be employed for other relevant applications.

\subsubsection{Handling Packet Errors and Erasures in the Network}
During a node-repair or a data-reconstruction operation, due to noise in the network, one or more packets downloaded may contain (arbitrary) errors. This is equivalent to a situation in the security scenario in which a set of nodes that are compromised to an active adversary may transmit corrupt packets. Our codes for on-demand security from active adversaries can alternatively be employed for handling packet-errors and erasures in the network.  The reader is referred to a conference version~\cite{ourISIT2012} of this paper for more details on this application.

The codes for security can correct for packet erasures as well. In this case, a node-repair (or data-reconstruction) operation guarding against $p$ errors and $p'$ erasures is carried out by downloading $\beta$ (or $\alpha$) symbols each from any arbitrary $(d+2p+p')$ (or $k+2p+p'$) nodes. An erasure of some $p'$ of these packets leaves us with data from some arbitrary subset of $(d+2p)$ (or $(k+2p)$) nodes, $p$ of which could be in error. This resultant scenario is identical to the setting of Section~\ref{sec:A_model}, which is addressed by our codes. The properties of on-demand security and optimality in the amount of download are also retained in the case of erasures.

\subsubsection{Reducing Latency of Degraded Reads in Data Centers}
The property of on-demand security can also aid in the reducing the latency of \textit{degraded reads} in data centers. A degraded read (also called \textit{online repair}) is an operation which is executed when a request comes in for data stored in a node that is busy or has failed. Under this operation, the request is met by downloading and recovering the requested data from the remaining nodes via a repair operation. Degraded reads are typically latency-sensitive, i.e., they must be served quickly. 

If the underlying code provides on-demand security, then it allows for the degraded read operation to be executed faster in the following manner. Let $f$ denote the busy/failed node that stores the requested data. Now, one could use the repair property of the erasure code to recover the required data from any $d$ other nodes. However, due to various sources of randomness in the system such as congestion at the nodes and delay in transmission, the replies from these $d$ nodes will typically arrive at different times. As a result, the net latency will be the maximum of the response-times of these $d$ nodes. On the other hand, a code providing on-demand security allows a reduction in latency by means of sending `redundant-requests'~\cite{vulimiri2012more,ananthanarayanan2012let,dean2013tail,shah2013redundant,joshi2014delay} as follows.  For some choice of a parameter $r~(d\leq r \leq n)$, connect to any $r$ nodes and ask each of these nodes to aid in the `repair' of node $f$. Since the code provides on-demand security, one can recover the data of node $f$ (and hence serve the request) from the data obtained from the \textit{first $d$ nodes that reply} (the requests to the remaining nodes are then canceled). This lowers the latency of serving this request to the response-time of the $d\supth$ fastest among the $r~(d\leq r \leq n)$ helper nodes.

\subsection{Notational Conventions} A vector will be treated as a column vector by default, and a row vector will be written as the transpose of the corresponding column vector. The transpose of a vector or a matrix will be denoted by a superscript $T$. The term `randomly drawn' will mean `drawn uniformly at random'. 

\section{Related Literature}\label{sec:literature}
\subsection{Codes for Distributed Storage}
The `regenerating codes' model introduced in~\cite{dimakis2010network} considers optimizing two important resources: the storage capacity required per node, and the repair-bandwidth. It was shown in~\cite{dimakis2010network} that there exists a tradeoff between these two resources, and lower bounds on their requirements were derived. Subsequent to the work of~\cite{dimakis2010network}, several explicit codes~\cite{ourProductMatrix,ourAllerton,ourITW,le2012exact,hu2013analysis,suh2011journal,tamo2011mds, papailiopoulos2013repairTransactions, cadambe2011permutation,cadambe2011polynomial,ourAllertonJournal} were constructed for the MSR and the MBR regimes of regenerating codes, many of which meet these bounds. Furthermore, it was shown in~\cite{ourAllertonJournal,tian2013rate,sasidharan2013improved} that the bounds are loose at essentially all points in the interior of the tradeoff curve.

A class of explicit codes that are optimal in terms of the storage and bandwidth requirements are the product-matrix codes proposed in~\cite{ourProductMatrix}. 
The results of this paper are based on the product-matrix codes, and exploit certain unique features of the underlying product-matrix framework. A detailed description of the product-matrix codes is provided in Section~\ref{sec:PM}.

The requirement of security in the presence of repair dynamics was first considered in~\cite{pawar2011securing}. In~\cite{pawar2011securing}, the authors provide lower bounds on the storage and bandwidth requirements under such a setting. Secure codes for the MBR regime with $d=n-1$ are also provided, which are based on the repair-by-transfer code of~\cite{ourAllerton,ourAllertonJournal}. The codes of~\cite{pawar2011securing} providing security from active adversaries, also for the MBR regime with $d=n-1$, allow propagation of the errors during the repair operation, and treat the resulting errors only at the reconstruction stage. Such a property may not be desirable in practice since the system administrators may not approve of allowing errors to linger around and propagate through the system. The secure codes presented in the present paper are applicable to any choice of the system parameters $[n,k,d]$ in the MBR regime, and any $[n,k,d \geq 2k-2]$ in the MSR regime. Furthermore, our codes ensure the correction of errors during every individual repair and the reconstruction operation. While this additional requirement clearly makes the system more practical, we show that surprisingly, this more stringent requirement does \textit{not} necessitate any additional storage or bandwidth requirements.




Subsequent to the initial presentation of our codes in~\cite{ourGlobecom2011}, there have been several other works on information theoretic security in distributed storage systems with repair considerations~(\cite{silberstein2012adversarial,rawat2014optimal,goparaju2013data,tandon2013towards, ernvall2013capacity}). In~\cite{rawat2014optimal}, the authors consider security from passive eavesdroppers and present outer bounds for the MSR setting that are tighter than those proposed in~\cite{pawar2011securing}. The authors also present a secure MSR code construction for $d=n-1$ which employs Zigzag codes~\cite{tamo2013zigzag} along with a maximum rank distance code. These codes are optimal for $d=n-1, \, \, m\leq 2$.  The secrecy capacity at the MSR regime for $m > 2$ remains open. 
In~\cite{goparaju2013data}, the authors provide a tighter upper bound on the file size that can be stored in a system secure from passive eavesdroppers for codes employing linear encoding and decoding at the MSR point. The code construction provided in~\cite{rawat2014optimal} meets the upper bound for linear codes provided in~\cite{goparaju2013data} for $d=n-1$. 

Security from active adversaries is considered in~\cite{silberstein2012adversarial}. A different adversarial model, where an active adversary can replace the content of an affected node only once, is considered and bounds and achievable schemes for this setting are provided. The paper also considers the MSR setting providing security from an active adversary who can replace the contents of affected nodes an unbounded number of times, and provides schemes that are optimal for a specific choice of the parameters. In \cite{tandon2013towards}, the authors provide information theoretic upper bounds for the cases $n \leq 4$ and $k=d=n-1$. For these parameters, it is shown that MBR is the one and only regime for which the resource requirements in the system do not increase when, in addition to security from an eavesdropper being able to read data stored in nodes, security from an eavesdropper who can also tap on to the data downloaded during repair is desired. Security in distributed storage systems with heterogeneous storage nodes is  considered in~\cite{ernvall2013capacity}.


In~\cite{han2011exact}, the authors deal with byzantine fault tolerance by employing product-matrix codes~\cite{ourProductMatrix}. They use a cyclic redundancy check (CRC) to check the integrity of data during repair and reconstruction, and a feedback scheme to iteratively correct them. However, CRC based schemes are not applicable in the present setting of information-theoretic security since the CRC may also be corrupted by the adversary. The present paper takes a more fundamental look at the problem of handling corruptions in regenerating codes from an information-theoretic perspective. 

In~\cite{oggier2011byzantine}, the authors derive bounds to determine the secure capacity in a `cooperative-repair' setting~\cite{shum2011exact}. The bounds of~\cite{oggier2011byzantine} show that such an attempt to cooperatively repair may adversely affect the system in the presence of malicious adversaries. In~\cite{dikaliotis2010security} authors study the security of distributed storage systems in the presence of a trusted verifier. In~\cite{kosut2013polytope}, authors propose using polytope codes to address the issue of security against active adversaries.  

\subsection{Shamir's secret sharing}
A possible method to ensure information-theoretic security from passive eavesdroppers is to employ Shamir's secret sharing scheme~\cite{shamir1979share}. Under Shamir's secret sharing scheme, the data is encoded and stored in a set of $n$ nodes such that the entire data can be recovered from any $k$ nodes, while access to data in any $(k-1)$ or fewer nodes provides zero information about the data. Now, during repair of a failed node, this scheme requires a download of the entire data to a central location, following which the data in the replacement node is re-encoded. Thus, as in the case of classical erasure codes, the repair operations are inefficient, mandating significant network resources. Furthermore, this central location represents a single point of failure, and security in such a system can be compromised by an eavesdropper who gains access to this location. These critical issues thus necessitate investigation of alternative solutions that account for the routinely performed failure-handling tasks.

\subsection{Secure Network Coding} 
The literature on secure network coding (e.g.,~\cite{cai2011secure,feldman2004capacity,yao2010network,ngai2009secure}) primarily considers a \textit{multicast} setting in which there is a single source of data and every destination is interested is obtaining \textit{all} the data transmitted by the source. Furthermore, with respect to security from passive eavesdroppers in the multicast setting, only the scenarios where the eavesdropper can access subsets of \textit{links} is well understood in the literature.  The problem of secure distributed storage considered in this paper requires handling the case when nodes are compromised. 
The problem of node-compromise is typically treated as a case of link-compromise by assuming that the eavesdropper gains access to all links that are incident upon the compromised nodes. 
In our problem, since a node may be repaired by connecting to \textit{any} $d$ nodes, schemes following such an approach cannot have a non-zero rate of transmission in general.

Even in the case of active adversaries, the problem of handling compromise of nodes is much harder~\cite{kosut2009nonlinear} than handling compromise of links. Furthermore, our problem falls into the harder setting of non-multicast, whereas the literature on secure network coding primarily considers the multicast setting. The works~\cite{koetter2008coding,silva2008rank} present results showing guarantees of error correction under random linear network coding under the condition that the subspace obtained at the receiver has a sufficient intersection with the transmitted subspace, a condition that is not guaranteed in our setting. In~\cite{jaggi2007resilient}, the authors propose schemes to transmit a message equal to the difference between the largest message that can be sent in the absence of secrecy requirements and a bound on the number of compromised links. In our problem, this difference is almost always $0$ or smaller since a compromised node may assist in any number of repairs of any other node. The results of the present paper, on the other hand, establish the capacity of a class of non-multicast networks in the presence of active adversaries or passive eavesdroppers having the ability to compromise nodes. Interestingly, these capacity-achieving codes that we propose in this paper are linear, deterministic and explicit.

The problem of providing information-theoretic secrecy in distributed storage systems is also related to the Wiretap Channel II~\cite{ozarow1985wire} where an eavesdropper, who is listening to any arbitrary subset of symbols (of fixed size) being transmitted over a noiseless point-to-point channel, should obtain no information about the original message. While schemes providing secrecy in a distributed storage system  with only the reconstruction requirement would follow directly from~\cite{ozarow1985wire}, the requirement of addressing node-repair makes the problem non-trivial. 

\section{The Product-Matrix Framework}\label{sec:PM}
The secure codes presented in this paper are based on the \textit{product-matrix} codes~\cite{ourProductMatrix}, and this section describes their underlying framework. These codes operate at both extremities of the storage-bandwidth tradeoff for the parameters:
\begin{enumerate}
\item MBR: all parameters $[n, \ k, \ d]$, and
\item MSR: all parameters $[n, \ k, \ d \geq 2k-2]$~.
\end{enumerate}

Observe from~\eqref{eq:MBR_parameters_initial} and~\eqref{eq:MSR_parameters_initial} that for each of these cases, the values of $B$ and $\alpha$ are multiples of $\beta$. It follows that given an optimal code for $\beta=1$, optimal codes for any higher value of $\beta$ (say, $\beta_0$) can be obtained by simply concatenating the $\beta=1$ code $\beta_0$ number of times. This process is known as \textit{striping}. Thus in this paper, without loss of generality, we consider only the case of $\beta=1$.

\begin{figure*}[t]
\centering
\subfloat[Encoding]{
\includegraphics[height=2.0in]{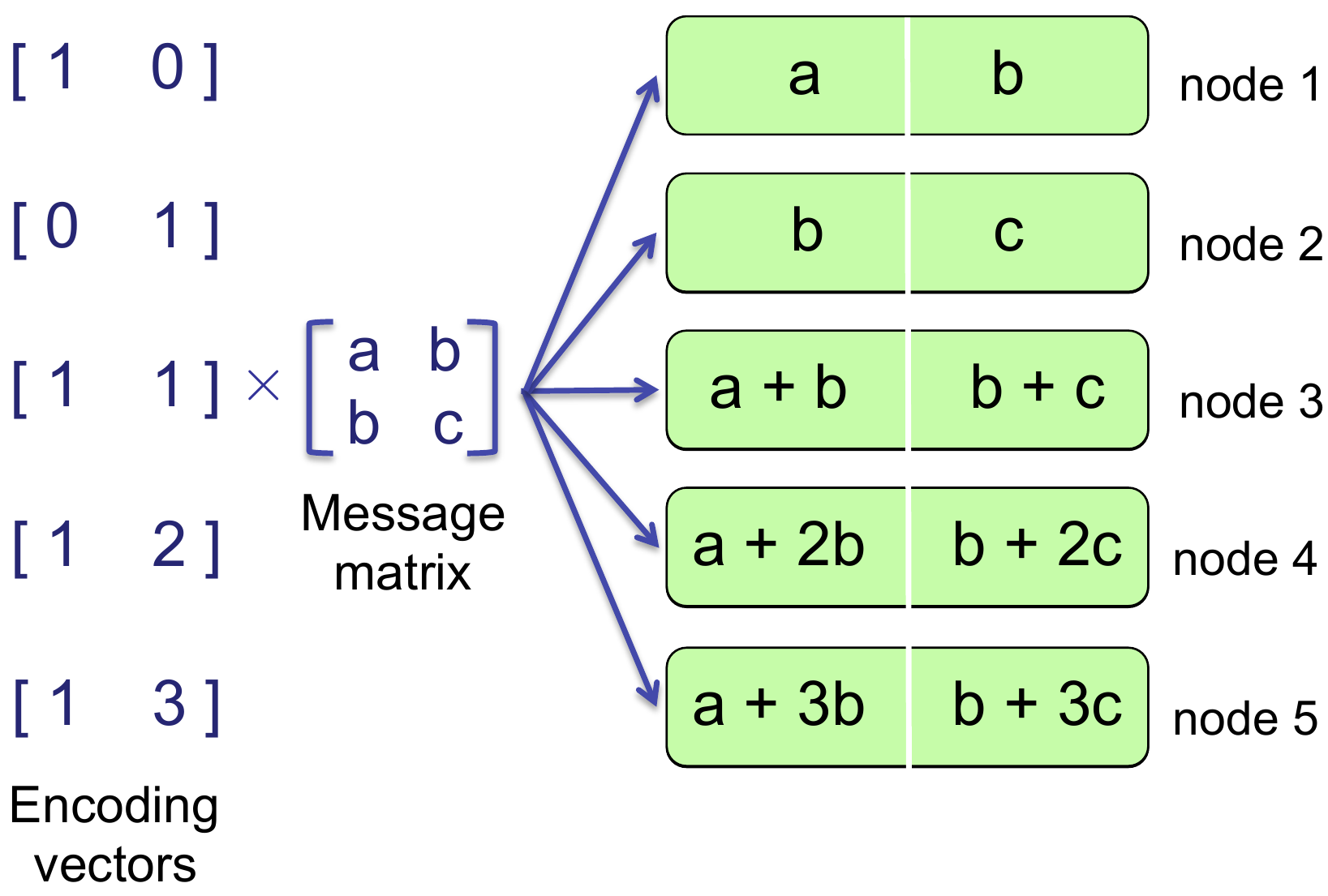}
}
\subfloat[Repair]{
\raisebox{.31in}{
\includegraphics[height=1.7in]{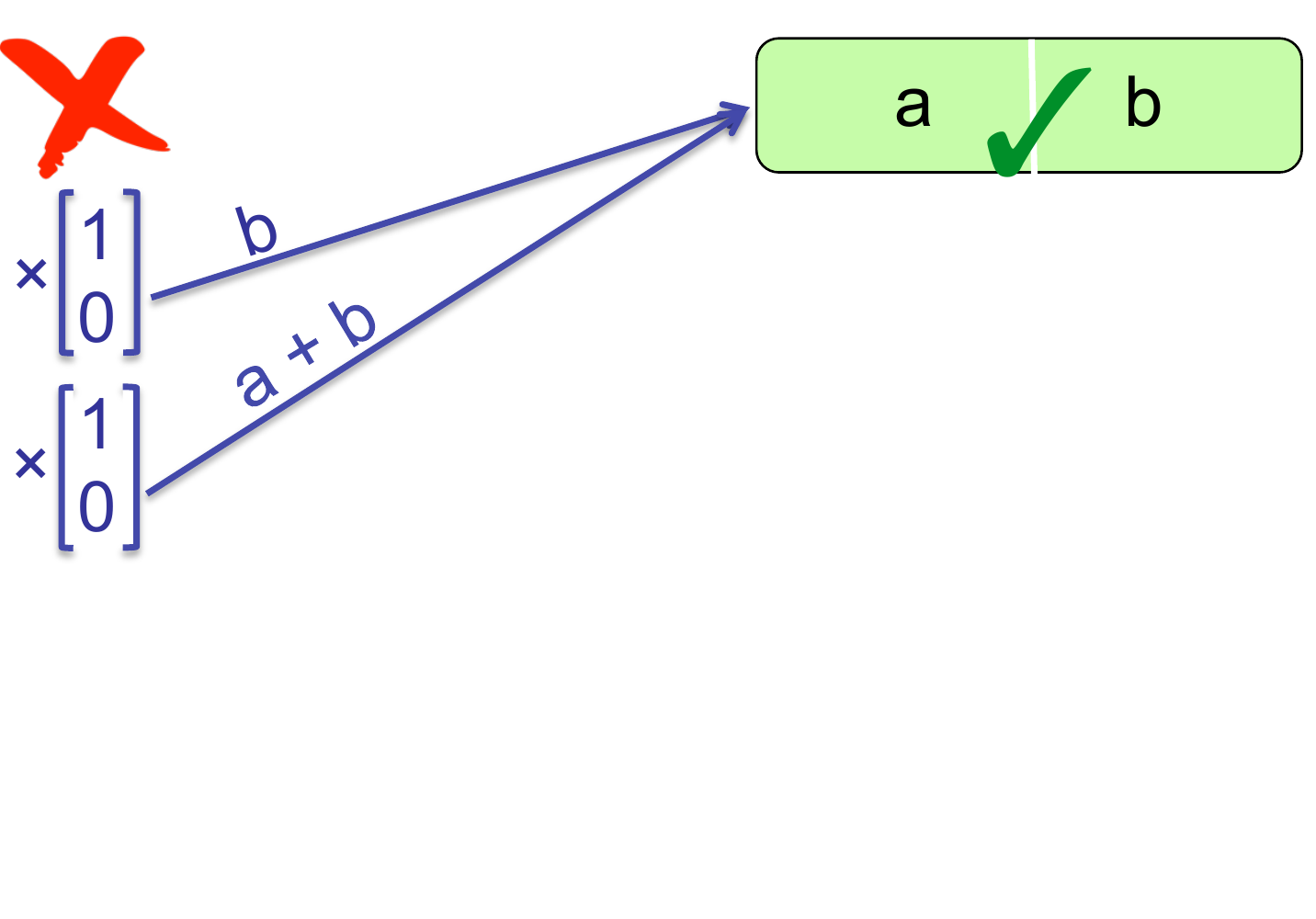}
}}
\caption{An example illustrating the product-matrix framework. The encoding of an MBR code for parameters $[n=5,\ k=2,\ d=2]$, $(B=3,\ \alpha=2,\ \beta=1)$ and $\ell=m=0$ (i.e., no requirement of security from passive eavesdroppers) is depicted. The field of operation is $\mathbb{F}_5$. Also depicted is the repair of the first node with $p=0$ (i.e., no requirement of security from active adversaries).
}
\label{fig:PM_MBR_encoding_example}
\end{figure*}

Product-matrix codes are represented in terms of an $(n \times \alpha)$ code matrix \beq C= \Psi M~, \eeq where
\begin{itemize}
\item the `\textit{code matrix}' $C$ is an $(n \times \alpha)$ matrix with its $i\supth$ row comprising the $\alpha$ symbols stored in node $i$~($1\leq i \leq n$),
\item the `\textit{message matrix}' $M$ is a $(d \times \alpha)$ matrix, whose elements comprise all the message symbols, along with some random symbols, arranged in a specific (and possibly redundant) manner, and 
\item the `\textit{encoding matrix}' $\Psi$ is a fixed $(n \times d)$ matrix. Denoting the $i^{th}$ row of $\Psi$ by ${\psi}_{i}^{T}$, the $\alpha$ symbols stored in node $i$ are \[ {\psi}_{i}^{T} M~.\] The $d$ length vector ${\psi}_{i}^T$ is termed as the \textit{encoding vector} of node $i$.
\end{itemize}
Thus each node stores the inner product of its encoding vector with the message matrix. The specific structures of the matrices $M$ and $\Psi$ vary with the choice of the operating regime (MBR or MSR) and the level of security required, and their design is described in the subsequent sections. 
Fig.~\ref{fig:PM_MBR_encoding_example} depicts an example of a code constructed under the product-matrix framework.



\subsubsection*{Systematic Codes}\label{sec:model_systematc}
A systematic code is one wherein some set of $k$ nodes together store all the $B$ message symbols in an uncoded form. This set of $k$ nodes are called systematic nodes. In several applications of interest, it may be desired to have the code in a systematic form, since decoding of the data is not required whenever the data-collector connects to the $k$ systematic nodes.

The codes for security from active adversaries that presented in this paper can be constructed in both systematic or non-systematic forms. On the other hand, a code operating under the requirement of information-theoretic security from passive eavesdroppers \textit{cannot} be systematic. This is because, if any part of the message is stored in an uncoded form in any node, an eavesdropper tapping onto that node would obtain a non-zero amount of information about the message.

The following sections employ the product-matrix framework to construct secure MBR and MSR codes.

\section{MBR codes Secure from Active Adversaries}\label{sec:A_MBR}
MBR codes achieve minimum possible download during repair: a replacement node downloads only what it stores, resulting in $d\beta = \alpha$. Recall from Section~\ref{sec:RC} (see~\eqref{eq:MBR_parameters_initial}) that in the absence of security requirements, an MBR code must satisfy
\beq \alpha=\frac{B}{k} + \frac{(k-1)\beta}{2},~~~d\beta=\alpha~. \label{eq:MBR_parameters}\eeq

In this section we present explicit constructions of optimal MBR codes for all parameter values $[n, \ k, \ d]$, providing security from active adversaries. As discussed in Section~\ref{sec:A_model}, our code constructions, in addition to being optimal, provide on-demand security. These codes are optimal in the sense that they achieve~\eqref{eq:cut_set_bound_secure_active_translated} with equality for every value of the desired security level $p$.

\subsection{Encoding algorithm}\label{sec:A_MBR_encoding}
The encoding procedure of the codes for security from active adversaries is identical to that of the product-matrix codes without security requirements~\cite{ourProductMatrix}, and is independent of the desired protection levels. 
This procedure is briefly described below for the sake of completeness. 

We apply the \textit{striping} procedure described in Section~\ref{sec:PM}, and construct codes for $\beta=1$. Identical encoding, repair and reconstruction operations are performed independently on each stripe of $\beta=1$. Setting $\beta=1$ in~\eqref{eq:MBR_parameters} gives
\beq \alpha=\frac{B}{k} + \frac{k-1}{2},~~~d=\alpha,~~~\beta=1~.\label{eq:MBR_parameters_striped} \eeq 
Recall from Section~\ref{sec:PM} that the product-matrix codes are represented in terms of an $(n \times \alpha)$ code matrix \beq C= \Psi M~, \label{eq:code_matrix_A_MBR}\eeq where the $i^{th}$ row of $C$ contains the $\alpha$ symbols stored in node $i$~$(1 \leq i \leq n)$. Thus the $i^{th}$ node stores $\psi_i^T M$ where $\psi_i^T$ is the $i^{th}$ row $\Psi$, and is termed the encoding vector of node $i$.

The encoding matrix $\Psi$ for the MBR code is of the form  
\beq \Psi = [\Phi ~~\Sigma]~,\label{eq:MBR_psi}\eeq 
where $\Phi$ is an $(n \times k)$ matrix, and $\Sigma$ is an $(n \times (d-k))$ matrix. The matrices $\Phi$ and $\Sigma$ are chosen such that: (i) any $k$ rows of $\Phi$ are linearly independent, and (ii) any $d$ rows of $\Psi$ are linearly independent. These requirements can be met, for example, by choosing $\Psi$ to be either a Cauchy or a Vandermonde matrix~\cite{cauchy}. Observe that these properties make $\Phi$ the generator matrix of an $[n, \ k]$ MDS code, and $\Psi$ the generator matrix of an $[n, \ d]$ MDS code. 
The choice of the finite field $\mathbb{F}_q$ is restricted (only) by the matrix $\Psi$: choosing $\Psi$ as a Vandermonde matrix permits any $q \geq n$.

The message matrix $M$ contains the $B$ message symbols arranged in a specific form. As mentioned previously, in our approach, the encoding mechanism is independent of the number of compromised nodes in the system. Hence the number of message symbols is equal to that in the regenerating codes setting with no security requirements~\eqref{eq:MBR_parameters_striped}: 
\beq B= kd - \frac{k(k-1)}{2}= k(d-k)+\frac{k(k+1)}{2}~.\label{eq:A_MBR_B}\eeq
The $(d \times d)$ message matrix is designed to be of the form 
\beq M = \left[\begin{array}{cc}S&\matT\\ \matT^T&0\end{array}\right],\label{eq:A_MBR_M}\eeq
where $S$ is a $(k \times k)$ \textit{symmetric} matrix, and $\matT$ is a $(k \times (d-k))$ matrix. The $B$ message symbols populate the $\frac{k(k+1)}{2}$ distinct entries of $S$ and the $k(d-k)$ entries of $\matT$. Observe that this specific structure makes the message matrix $M$ \textit{symmetric}.

In the MBR toy example provided in Fig.~\ref{fig:PM_MBR_encoding_example}, with $[n=5,\,k=2,\,d=2]$ and $(\beta=1,\,\alpha=2,\,B=3)$, the $(5 \times 2)$ encoding matrix $\Psi$ and the $(2 \times 2)$ message matrix $M$ take the values:
\[
\Psi = \begin{bmatrix}
1 & 0 \\
0 & 1 \\
1 & 1 \\
1 & 2 \\
1 & 3
\end{bmatrix}~~~,~M = \begin{bmatrix}
a & b \\
b & c
\end{bmatrix}~,
\]
with $a, \ b, \text{and } c$ as the three message symbols.

\begin{remark}[Systematic code] The PM-MBR code can be made systematic by choosing the encoding matrix $\Psi$ as
\beq \Psi = \left[
\begin{tabular}{>{$}c<{$} >{$}c<{$}}
I_k &  0 \\
\tilde{\Phi} & \tilde{\Sigma}
\end{tabular}
\right], \eeq where $I_k$ is the $(k \times k)$ identity matrix, $0$ is a $(k \times (d-k))$ zero matrix, $\tilde{\Phi}$ and $\tilde{\Sigma}$ are matrices of sizes $((n-k) \times k)$ and $((n-k) \times (d-k))$ respectively, such that $\left[\tilde{\Phi}~~~\tilde{\Sigma}\right]$ is a Cauchy matrix (or any matrix such that all of its submatrices are of full rank). The example in Fig.~\ref{fig:PM_MBR_encoding_example} depicts a systematic product-matrix MBR code.
\end{remark}

This completes the description of the encoding procedure. We now present the decoding algorithms that ensure security from active eavesdroppers.

\subsection{Algorithm for Secure Node Repair}\label{sec:A_MBR_repair}
The following theorem presents an explicit algorithm for node-repair in the presence of compromised nodes.


\begin{thm}[Secure node-repair]\label{thm:repair_A_MBR}
In the code presented, for any value of parameter $p$, the repair of a failed node can be secured from the compromise of up to $p$ nodes, by letting the replacement node download $\beta=1$ symbols each from $(d+2p)$ arbitrary nodes. The amount of data downloaded in this processes is minimum, thus establishing the secure capacity of such systems. The choice of parameter $p$ is arbitrary, and can be different for every instance of repair, thus providing on-demand security.
\end{thm}
\begin{IEEEproof}
Suppose during any instance of node-repair, it is desired to provide security from the compromise of up to $p$ nodes, for some choice of $p$. Let $f$ denote the failed node and $\psi_f$ be its encoding vector. The $\alpha~(=d)$ symbols stored in the failed node are \[ \psi_f^T M~.\]
The replacement node then connects to some arbitrary $(d+2p)$ nodes. The replacement node is required to recover these $d$ symbols by downloading $\beta=1$ symbol each from the $(d+2 p)$ nodes that it connects to, when at most $p$ of these nodes are compromised. Let $\mathcal{J}$ denote the set of these $(d+2 p)$ nodes. Under our repair algorithm, each of these $(d+2 p)$ nodes pass the inner-product of the $\alpha$ values stored in them with the encoding vector of the failed node $f$. That is, for every $j\in \mathcal{J}$, node $j$ passes the value of
\[ \psi_{j}^T M \psi_f ~. \]
This value may be corrupt if node $j$ is compromised to the adversary. 
Let $\Psi_\setJ$ denote the $((d + 2 p) \times d)$ submatrix of $\Psi$, with its $(d+2p)$ rows comprising the vectors $\lbrace \psi_j^T \rbrace_{j \in \mathcal{J}}$. Under this notation, one can see that the replacement node has access to the $(d+2 p)$ encoded symbols
\[ \Psi_\setJ M {\psi}_f ~,\]
of which at most $p$ symbols may be in error. This set of symbols can equivalently be viewed as an encoding of $M \psi_f$ by the generator matrix $\Psi_\setJ$. To complete the decoding process, we now call upon the MDS property of matrix $\Psi$. By construction, $\Psi$ is the generator matrix of an $[n, \ d]$ MDS code. Hence its submatrix $\Psi_\setJ$ is the generator matrix of a $[d+2p,\ d]$ MDS code. This implies that the code with $\Psi_\setJ$ as its generator matrix has a minimum distance of $(2p+1)$. Thus, when at most $p$ of the $(d+2p)$ symbols $\Psi_\setJ M {\psi}_f$ are corrupt, the replacement node can apply standard decoding algorithms for MDS codes and recover $M {\psi}_f$ correctly. Finally, since the message matrix $M$ is symmetric, the $d$ elements comprising $M {\psi}_f$ can be written as 
\[ (M {\psi}_f)^T = {\psi}_f^T M~, \]
which are precisely the $d$ values that were stored in the failed node. 

The repair algorithm described above required the replacement node to connect to some arbitrary $\Delta = (d+2p)$ nodes, and download $\beta=1$ symbol from each. The amount of storage and repair-bandwidth consumed meet~\eqref{eq:A_MBR_B} and~\eqref{eq:cut_set_bound_secure} with equality, thus establishing optimality.
\end{IEEEproof}

\begin{example}
Fig.~\ref{fig:example_adversary_MBR} depicts an example of this algorithm where repair of the first node is performed, while providing protection from up to $p=1$ compromised nodes. The encoding process of the code is illustrated in Fig.~\ref{fig:PM_MBR_encoding_example}. For the repair of node $1$ as shown in the figure, the replacement node connects to $\Delta = d+2p = 4$ other nodes and each of these nodes pass an inner-product of their data with $\psi_1 = [1~~0]$, i.e., pass their first symbol. The data $\{b,\ a+b,\ a+2b,\ a+3b\}$ is an MDS encoding of $\{a,\ b\}$ allowing for the correction of up to one arbitrary error.
\end{example}

\subsection{Algorithm for Secure Data Reconstruction}\label{sec:A_MBR_recon}
The following theorem presents an explicit algorithm for data reconstruction in the presence of compromised nodes. 

\begin{thm}[Secure data-reconstruction]
In the code presented, a data-collector can recover all the $B$ message symbols by downloading data stored in $(k+2p)$ arbitrary nodes, in the presence of up to $p$ compromised nodes. The choice of parameter $p$ is arbitrary, and can be different for every instance of data-reconstruction, thus providing on-demand security.
\end{thm}
\begin{IEEEproof}
Suppose during any instance of data-reconstruction, it is desired to provide protection from the compromise of up to $p$ nodes, for some choice of $p$. The data-collector then connects to $(k+2p)$ arbitrary nodes. Let $\mathcal{I}$ denote the set of $(k+2p)$ nodes to which the data-collector connects. Then, from every node $i\in \mathcal{I}$, the data collector downloads the $\alpha$ symbols
\[ {\psi}_{i}^T M~, \]
which could be corrupt if node $i$ is compromised. Let $\Psi_\setI$ denote the $((k + 2p) \times d)$ submatrix of $\Psi$, with its $(k+2p)$ rows comprising $\lbrace \psi_i^T \rbrace_{i \in \mathcal{I}}$. From the structure of the $\Psi$ (see~\eqref{eq:MBR_psi}) we have
\beq \Psi_{\setI} = \left[ \begin{tabular}{>{$}c<{$}>{$}c<{$}} \Phi_{\setI} & \Sigma_{\setI} \end{tabular} \right]~, \eeq
where $\Phi_\setI$ is a $((k+2p)\times k)$ matrix and $\Sigma_\setI$ is a $((k+2p)\times (d-k))$ matrix. The $(k+2p)$ rows of  $\Phi_\setI$ and  $\Sigma_\setI$ respectively are comprised of $(k+2p)$ of the rows of $\Phi$ and $\Sigma$ (indexed by $\mathcal{I}$). From the specific structure~\eqref{eq:A_MBR_M} of the message matrix $M$, we see that the data-collector equivalently has access to the encoded symbols  
\[ \Psi_\setI M = \left[ \begin{tabular}{>{$}c<{$}>{$}c<{$}} \Phi_\setI S  + \Sigma_\setI \matT^T  & \Phi_\setI \matT \end{tabular} \right].\]
Note that the $(k+2p)$ rows of the matrix $\Psi_\setI M$ are obtained from $(k+2p)$ different helper nodes. Thus, a compromise of up to $p$ of the helper nodes to an active adversary leads to at most $p$ of the rows of $\Psi_\setI M$ being corrupt.

We now call upon the MDS property of $\Phi$ to complete the reconstruction process. Recall that by construction, $\Phi$ is the generator matrix of an $[n, \ k]$ MDS code. Hence its submatrix $\Phi_\setI$ is the generator matrix of an $[k+2p,\ k]$ MDS code. It follows that a code with $\Phi_\setI$ as its generator matrix has a minimum distance of $(2p+1)$, and thus has the ability to correct up to $p$ arbitrary errors. Consider the set of symbols $\Phi_\setI \matT$ that the data collector has obtained. Observe that any column of the matrix $\Phi_\setI \matT$ is an encoding of the corresponding column of $\matT$ with $\Phi_\setI$ as the generator matrix. Thus, when at most $p$ of the nodes are compromised, the data-collector can correctly decode each column of $\Phi_\setI \matT$ separately using standard algorithms for decoding MDS codes. At the end of this process, the data-collector recovers the matrix $\matT$ correctly. The next step is to decode $S$ from the downloaded data. Since the value of $\matT$ is (correctly) known, the data-collector can subtract $\Sigma_\setI \matT^T$ from $(\Phi_\setI S  + \Sigma_\setI \matT^T)$ to recover $\Phi_\setI S$. Again, the matrix $\Phi_ \setI S$ is an encoding of the columns of $S$ by the generator matrix $\Phi_\setI$, which allows for the decoding of $S$ in a manner identical to the decoding of $\matT$ even in the presence of upto $p$ corrupt rows.

In this setting, one can verify that the amounts of storage and repair-bandwidth consumed meet~\eqref{eq:cut_set_bound_secure_active} with equality, thus establishing optimality.
\end{IEEEproof}

\section{MBR codes for Security from Passive Eavesdroppers}\label{sec:E_MBR}
In this section, we present explicit constructions of MBR codes secure from passive eavesdroppers that support all values of $[n, \ k, \ d]$ and all $\{\ell,m\}$. 
As discussed previously, unlike our codes for security from active adversaries, the codes for security from passive eavesdroppers do not provide on-demand security. As a result, the desired level $\{\ell,\,m\}$ of security must be chosen apriori at the time of encoding. 
The MBR regime is governed by the relation $\alpha = d\beta$, and substituting this in~\eqref{eq:cut_set_bound_secure} gives an upper bound on the amount of data $B^*$ that can be stored in a $\{\ell,\ m\}$-secure MBR code as
\beq B^* \leq \left(k d  - \frac{k(k-1)}{2} \right)\beta - \left(l d - \frac{\ell(\ell-1)}{2} \right)\beta~.\label{eq:mbr_secret_cutset_ineq}\eeq
The codes presented here achieve this bound with equality
\beq B^* = \left(k d  - \frac{k(k-1)}{2} \right)\beta - \left(l d - \frac{\ell(\ell-1)}{2} \right)\beta~, \label{eq:mbr_secret_cutset}\eeq
thus establishing this as the secure capacity under the MBR regime for all values of the parameters $[n,\ k, \ d]$ and $\{\ell,\ m\}$.

When any parameter takes different values in the secure and non-secure versions, we will denote the parameter associated to a code providing security with a superscript asterisk $(^*)$ as done in~\eqref{eq:mbr_secret_cutset_ineq} and~\eqref{eq:mbr_secret_cutset}.

\subsection{Encoding}\label{sec:E_MBR_encoding}
We will now construct an $\{\ell,\ m\}$-secure MBR code satisfying~\eqref{eq:mbr_secret_cutset}. 
We apply the striping procedure described in Section~\ref{sec:PM}, and (without loss of generality) construct codes for the case $\beta=1$. Setting $\beta=1$ in~\eqref{eq:mbr_secret_cutset} gives
\[ B^*=\left(k d  - \frac{k(k-1)}{2} \right) - \left(l d - \frac{\ell(\ell-1)}{2}\right),~~~\alpha=d,~~~\beta=1~.\]

The product-matrix codes for security from eavesdroppers possess a structure identical to those without security requirements. The difference between the two codes is that in the absence of security requirements, the elements of the message matrix $M$ consist of the message symbols, whereas in the situation when security is needed, a specific set of elements of the message matrix are replaced by random symbols. 
Thus, to construct an $\{\ell,\ m\}$-secure code for any $[n,\ k,\ d]$, first consider an $[n,\ k,\ d]$ product-matrix MSR code in the absence of security requirements, as constructed in Section~\ref{sec:A_MBR_encoding}. Denote the this code by $\mathcal{C}$, and further, let $\mathcal{C}^*$ denote the $\{\ell,\ m\}$-secure code (which we will construct below). The code $\mathcal{C}$ is described by an $(n\times \alpha)$ code matrix $C = \Psi M$ as defined in~\eqref{eq:code_matrix_A_MBR}, with the $(d \times d)$ matrix $M$ comprising $B = kd-\frac{k(k-1)}{2}$ message values~\eqref{eq:MBR_parameters_striped}. Following the product-matrix framework, the code $\mathcal{C}^*$ is also described by an $(n \times \alpha)$ code matrix $C^*$ of the form $C^* = \Psi^* M^*$. The matrices $\Psi^*$ and $M^*$ for code $\mathcal{C}^*$ are obtained by modifying the matrices $\Psi$ and $M$ of code $\mathcal{C}$ in a manner described below. 

The $(n \times d)$ encoding matrix $\Psi^*$ is required to satisfy the following property in addition to those required by $\Psi$: when restricted to the first $\ell$ columns, any $\ell$ rows are linearly independent. The choice of $\Psi^*$ as a Cauchy or a Vandermonde matrix satisfies this property as well.

\begin{figure}[t]
\centering
\includegraphics[width=.4\textwidth]{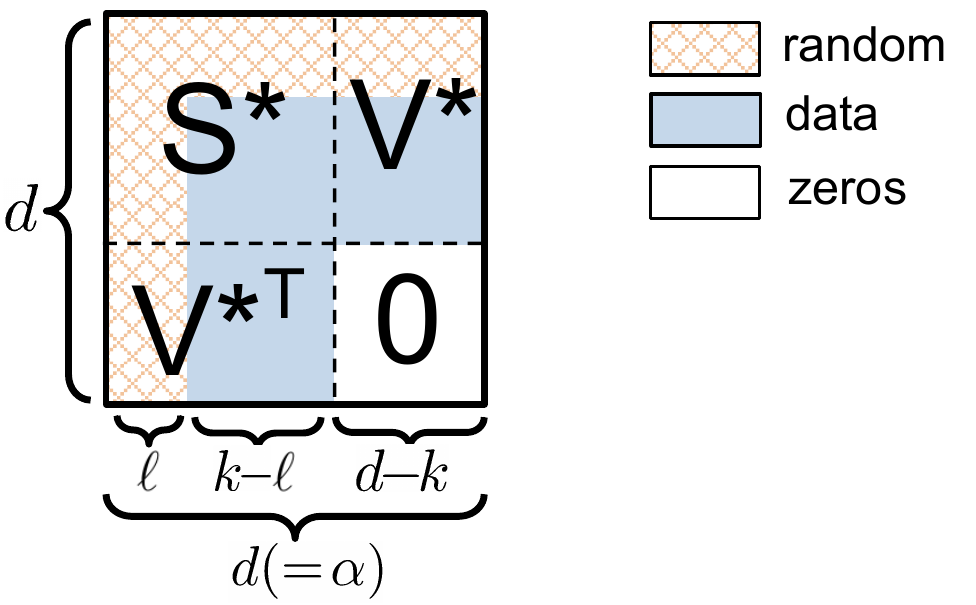}
\caption{Structure of the $(d \times d)$ message matrix $M^*$ for the $\{\ell,\ m\}$-secure MBR code. The matrix is symmetric and comprises of blocks of data, random symbols, and zeros. The code is given by $\mathcal{C}^* = \Psi^* M^*$ where $\Psi^*$ is the encoding matrix.}
\label{fig:eaves_M_MBR}
\end{figure}

The $(d \times d)$ message matrix $M^*$ is obtained by replacing the
\beq R = B - B^* = \ell d - \frac{\ell(\ell-1)}{2} \label{eq:mbr_number_random} \eeq message symbols in the first $\ell$ rows (and hence first $\ell$ columns) of the symmetric matrix $M$ by $R$ random symbols. Each random symbol is drawn independently and uniformly from $\mathbb{F}_q$. The structure of the resulting message matrix is illustrated in Fig.~\ref{fig:eaves_M_MBR}.

Finally, node $i~(1\leq i \leq n)$ stores the $i^{th}$ row of the $(n\times \alpha)$ code matrix $C^* = \Psi^* M^*$.

The following example illustrates the encoding procedure.
\begin{example}
Consider the code depicted in Fig.~\ref{fig:example_eaves_MBR} providing security from passive eavesdroppers. This code operates in the MBR regime with parameters $n=\nval$, $k=2$ and $d=2$, and provides security against eavesdroppers who may gain access to $(\ell=1,\ m=1)$ nodes. Let the alphabet of operation be $\mathbb{F}_{\nval}$ and let $\alpha=2$ and $\beta=1$. From~\eqref{eq:mbr_secret_cutset} we get that the maximum size of the message that can be stored securely in this system is $B^*=1$ symbol (we could have stored $B=3$ symbols in the absence of security requirements). Let us denote this solitary message symbol as $a$. Following~\eqref{eq:mbr_number_random}, we draw $R=2$ symbols $r_1$ and $r_2$ uniformly at random from $\mathbb{F}_{\nval}$. The encoding vector of any node $i~(1\leq i \leq \nval)$ under this code is $\psi_i^T=[1~~i]$. The message matrix is \[M^*=\left[ \begin{tabular}{cc} $r_1$&$r_2$\\$r_2$&$a$\end{tabular}\right]~.\] Each node $i$ stores $\psi_i^T M^*$ as shown in Fig.~\ref{fig:example_eaves_MBR}.
\end{example}

\subsection{Reconstruction, Repair and Security}
The following theorems prove the properties of reconstruction, repair and security.

\begin{thm}[Data-reconstruction and node-repair]\label{thm:E_MBR_recon_repair}
In code $\mathcal{C}^*$, a data-collector can recover all the $B^*$ message symbols by downloading the data stored in any $k$ nodes, and a failed node can be repaired by downloading $\beta=1$ symbol each from any $d$ remaining nodes.
\end{thm}
\begin{IEEEproof}
Treating the random symbols also as message symbols, code $\mathcal{C}^*$ becomes identical to $\mathcal{C}$. Thus reconstruction and repair in $\mathcal{C}^*$ are identical to that in $\mathcal{C}$. \end{IEEEproof}

Note that repair and reconstruction can be carried out ensuring security from corruption by active adversaries by employing the explicit algorithms of Section~\ref{sec:A_MBR_repair} and Section~\ref{sec:A_MBR_recon}.

\begin{thm}[Information-theoretic security]\label{thm:E_MBR_security}
The code $\mathcal{C}^*$ ensures $\{\ell,\ m\}$-security, i.e., an eavesdropper having access to the data stored in up to $\ell$ nodes, and to all the data passed to up to $m$ of these nodes during one or more of their repair operations, gets no information about the message.
\end{thm}
\begin{IEEEproof}
The complete proof is provided in the appendix, and we provide an outline here. Let $\mathcal{U}$, $\mathcal{R}$ and $\mathcal{E}$ respectively denote random variables corresponding to the message, the random symbols inserted at the encoding stage, and the data available to the eavesdropper respectively. The proof proceeds in three steps: first it is shown that $H(\mathcal{R}|\mathcal{E},\mathcal{U})=0$. The second step is to prove $H(\mathcal{E})\leq R$. The final step shows that these two conditions suffice to guarantee complete security, i.e., $I(\mathcal{U};\ \mathcal{E})=0$.
\end{IEEEproof}

\section{MSR Codes Secure from Active Adversaries}\label{sec:A_MSR}


MSR codes use the minimum possible storage at each node. The data-reconstruction property necessitates that the data from any $k$ nodes suffice to reconstruct all the $B$ message symbols. As a result, each node must necessarily store at least a fraction $\frac{1}{k}$ of the entire message. Hence for an MSR code we have $\alpha = \frac{B}{k}$. To meet the bound~\eqref{eq:cut_set_bound} with equality (in absence of errors/erasures), an MSR code must satisfy
\begin{subequations}\begin{equation} B = k\alpha \label{eq:MSR_parameters_B} \end{equation} 
\begin{equation} d\beta = \alpha + (k -1)\beta~\label{eq:MSR_parameters_beta}.\end{equation} \label{eq:MSR_parameters}\end{subequations}

In this section we present explicit constructions of optimal MSR codes for all parameter values $[n, \ k, \ d\geq 2k-2]$ providing on-demand security from active adversaries. We employ the striping procedure described in Section~\ref{sec:PM}, and without loss of generality construct optimal codes for $\beta=1$.

Since our codes are required to provide on-demand security, the encoding procedure is independent of the levels of security desired and is identical to the case when no security is required.

We first construct the code for the case $d=2k-2$ and then use the \textit{shortening} technique of~\cite{ourITW,ourProductMatrix} to obtain codes for all $d \geq 2k-2$. As we will see subsequently, the decoding algorithms and the security properties of the code for $d > 2k-2$ follow directly from the corresponding properties of the $d=2k-2$ code.

\subsection{Encoding algorithm for $d=2k-2$} \label{sec:encoding_A_MSR}
In the MSR regime with $d=2k-2$ and $\beta=1$, from~\eqref{eq:MSR_parameters} we have
\begin{subequations} \bea k &=&  \alpha+1~, \label{eq:alpha_k_minus_1} \\
d&=& 2\alpha~, \label{eq:d_2alpha}\\
B &=& \alpha (\alpha+1)~.\label{eq:B_alpha_alphap1}\eea\end{subequations}

The product-matrix MSR code is described by an $(n \times \alpha)$ code matrix $C$ of the form $C=\Psi M$, with an $(n \times d)$ encoding matrix $\Psi$ and an $(d \times \alpha)$ message matrix $M$. The MSR code differs from the product-matrix MBR code in the specific design of the matrices $\Psi$ and $M$.

Under the MSR regime, the $(n \times d)$ encoding matrix $\Psi$ is of the form 
\beq \Psi = [\Phi ~~\Lambda \Phi]~, \label{eq:MSR_psi}\eeq 
where $\Phi$ is an $(n \times \alpha)$ matrix and $\Lambda$ is an $(n \times n)$ \textit{diagonal} matrix. The matrices $\Phi$ and $\Lambda$ are chosen such that: (i) any $\alpha$ rows of $\Phi$ are linearly independent, (ii) any $d$ rows of $\Psi$ are linearly independent, and (iii) the diagonal elements of $\Lambda$ are all distinct. These requirements can be met, for example, by choosing $\Psi$ to be a Vandermonde matrix with elements chosen carefully to satisfy the third condition: the $i\supth$ $(1\leq i \leq n)$ row of $\Psi$, $\psi_i^T = [1~~x_i~~x_i^2~\cdots~x_i^{d-1}]$, is chosen such that $x_{i_1}^\alpha \neq x_{i_2}^\alpha~~\forall~1\leq i_1 < i_2 \leq n$. Observe that these properties make $\Phi$ the generator matrix of an $[n, \ \alpha]$ MDS code, and $\Psi$ the generator matrix of an $[n, \ d]$ MDS code.

The choice of the matrix $\Psi$ poses the only restriction to the choice of the finite field $\mathbb{F}_q$. For instance, choosing $\Psi$ as a Vandermonde matrix (in the manner described above) permits any $q \geq n^2$.

We will now specify the design of the message matrix $M$. From~\eqref{eq:d_2alpha} and~\eqref{eq:B_alpha_alphap1}, we have
$d=2\alpha$ and $B = \alpha (\alpha+1)$ respectively. The $(d \times \alpha)$ message matrix $M$ is constructed as 
\beq M = \begin{bmatrix} S_1 \\ S_2\end{bmatrix}~, \label{eq:MSR_M_structure}\eeq 
where $S_1$ and $S_2$ are $(\alpha \times \alpha)$ symmetric matrices. The matrices $S_1$ and $S_2$ together have precisely $\alpha(\alpha+1)$ distinct entries, which are now populated by the $B=\alpha (\alpha+1)$ message symbols. Thus, under this encoding mechanism, node $i~(1 \leq i \leq n)$, stores the $\alpha$ symbols
\beq {\psi}_i^T M = {\phi}_i^T S_1+ \lambda_{i} {\phi}_i^T S_2~.\nonumber \eeq

This encoding procedure is illustrated below with an example. This example will subsequently be reused to illustrate secure node-repair.
\begin{example}\label{ex:MSR_encoding_nonsystematic}
Let us construct an MSR code for $[n=7, \ k=3, \ d=4]$. With $\beta=1$, the associated parameters are $(\alpha=2, \ B=6)$. Note that the chosen parameters $k$ and $d$ satisfy the relation $d=2k-2$. Let us operate over the finite field $\mathbb{F}_{13}$, and choose the encoding matrix $\Psi$ to be a Vandermonde matrix. Further, let us denote the $6$ message symbols as $\lbrace a, \ b,\ c,\ d,\ e,\ f \rbrace$. The matrices $\Psi$ and $M$ take values
\[
\Psi = \begin{bmatrix}
1&0&0&0\\
1 & 1&1&1 \\
1&3&9&1 \\
1&2&4&8\\
1&6&10&8\\
1&5&12&8\\
1&4&3&12
\end{bmatrix}
~~~~,~~
M = \begin{bmatrix}
a & b\\
b & c \\
d & e \\
e & f
\end{bmatrix}~.
\]
The code is given by $C=\Psi M$. The resultant code is depicted in Fig.~\ref{fig:adv_msr_nonsys}.
\end{example}

\begin{figure*}
\begin{center}
\renewcommand{\arraystretch}{\codeTableRowspace}
\begin{tabular}{|@{\colortablecell}c|ccccccc|ccccccc|}
\hline
\textrm{Node} & \multicolumn{14}{@{\colortablecell}c|}{Data Stored}\\
\hline
Node 1& a & & & & & & &  b & & & & & &\\
Node 2& a &+& b &+& d &+& e	&  b& +& c& +& e& +& f\\
Node 3& a &+& 3b &+& 9d &+& e	&  b& +& 3c &+ &9e& +& f\\
Node 4& a &+& 2b& +& 4d &+ &8e	&  b &+& 2c &+& 4e &+& 8f\\
Node 5& a &+& 6b &+& 10d &+& 8e & b &+& 6c &+& 10e &+& 8f	\\
Node 6& a &+& 5b &+& 12d &+& 8e &  b &+& 5c &+& 12e &+& 8f\\
Node 7& a &+& 4b &+& 3d &+& 12e & b &+& 4c &+& 3e &+& 12f	\\
\hline
\end{tabular}
\end{center}
\caption{An MSR code for $[n=7,\ k=3,\ d=4]$. The message is $\{a,b,c,d,e,f\}$ and the finite field of operation is $\mathbb{F}_{13}$. \advMSR{7}{4}{~3}{6}}
\label{fig:adv_msr_nonsys}
\end{figure*}

\begin{remark}[Systematic code]
Although the encoding mechanism described above does not directly result in a systematic code, it can be converted easily to systematic form by following a procedure that will be described subsequently in Section~\ref{sec:MSR_systematic}. 
\end{remark}

This completes the description of the encoding procedure. We now move on to present decoding algorithms for node-repair and data-reconstruction with on-demand security.

\subsection{Algorithm for Secure Node Repair}
The following theorem presents an explicit algorithm for node-repair that ensures security in the presence of compromise of nodes to an active adversary.
\begin{thm}[Secure node-repair]
In the code presented, for any desired value of $p$, the repair of a failed node can be secured from the compromise of up to $p$ nodes, by letting the replacement node download $\beta=1$ values each from $(d+2p)$ arbitrary nodes. The amount of data downloaded in this processes is minimum, thus establishing the secrecy capacity of such systems. The choice of parameter $p$ is arbitrary, and can be different for every instance of repair, thus providing on-demand security.
\end{thm}
\begin{IEEEproof}
Suppose during any instance of node-repair, it is desired to provide protection from the compromise of up to $p$ nodes, for some choice of $p$. Let $f$ denote the failed node and $\psi_f  = [\phi_f ~~\lambda_f\phi_f]$ be its encoding vector. The $\alpha~(=d)$ symbols stored in the failed node are
\beq {\psi}_f^T M = {\phi}_f^T S_1+ \lambda_{f} {\phi}_f^T S_2~. \label{eq:repair_alpha_A_MSR} \eeq
The replacement node connects to some $(d+2p)$ nodes. The replacement node is required to recover these $d$ symbols by downloading $\beta=1$ symbol each from the $(d+2 p)$ nodes that it connects to, when at most $p$ of these nodes are compromised. Let $\mathcal{J}$ denote the set of these $(d+2 p)$ nodes. Under our repair algorithm, each of these $(d+2 p)$ nodes pass the inner-product of the $\alpha$ values stored in them with the encoding vector $\psi_f$ of the failed node $f$. That is, for every $j\in \mathcal{J}$, node $j$ passes the value of
\[ \psi_{j}^T M \psi_f ~. \]
This value may be corrupt if node $j$ is compromised to the adversary. 

Let $\Psi_\setJ$ denote the $((d + 2p) \times d)$ submatrix of $\Psi$, with the rows of $\Psi_\setJ$ comprising the $(d+2p)$ rows of $\Psi$ that correspond to the $(d+2p)$ nodes in $\bigsetJ$.  Then the replacement node has access to the $(d+2p)$ values 
\[ \Psi_\setJ M {\phi}_f ~.\]
To simplify notation, let us define a quantity $\mu$ as \beq {\mu}=M {\phi}_f ~.\eeq
In terms of this notation, the $(d+2p)$ encoded symbols  downloaded by the replacement node can be written as $\Psi_\setJ {\mu}$. To decode in the presence of possible corruption, we will exploit the MDS property of the matrix $\Psi$. By construction, $\Psi$ is the generator matrix of an $[n, \ d]$ MDS code. Hence, $\Psi_\setJ$ is the generator matrix of a $[d+2p,\ d]$ MDS code. This implies that $\Psi_\setJ$ has a minimum distance of $(2p+1)$. The $(d+2p)$ symbols $\Psi_\setJ \mu$ downloaded by the replacement node are simply an encoding of $\mu$ under a code that has $\Psi_\setJ$ as its generator matrix. Hence, the replacement node can recover ${\mu}$ correctly by decoding the MDS code $\Psi_\setJ$ in the presence of up to $p$ corruptions. Thus, even in the presence of up to $p$ compromised nodes in the system, the replacement node correctly recovers
\bea {\mu} & = & M {\phi}_f \nonumber\\
&=& \left[ \begin{array}{c} S_1 \\ S_2\end{array} \right] {\phi}_f \\
&=& \left[ \begin{array}{c} S_1 {\phi}_f \\ S_2 {\phi}_f\end{array} \right]
\eea
Thus the replacement node has access to the correct values of $S_1 {\phi}_f$ and $S_2 {\phi}_f$. The symmetry of $S_1$ and $S_2$ allows for the computation of
\beq {\phi}_f^T S_1+ \lambda_{f} {\phi}_f^T S_2~ \nonumber\eeq
correctly, and this is precisely the collection of $\alpha$ symbols that were stored in the failed node. 
\end{IEEEproof}

The following example illustrates this repair algorithm.
\begin{example}
Consider the code described in Example~\ref{ex:MSR_encoding_nonsystematic} (Fig.~\ref{fig:adv_msr_nonsys}) for parameters $[n=7,\ k=3,\ d=4]$. Suppose node $2$ fails and needs to be repaired with a security from the compromise of any $p=1$ other node. From the construction described in Example~\ref{ex:MSR_encoding_nonsystematic}, we see that $\phi_f^T=[1~1]$. Under the secure repair algorithm described above, the replacement node connects to the $\Delta = d+2p=6$ remaining nodes. Each of these nodes takes an inner product of its data with the vector $[1~1]$, i.e., passes the sum of the two symbols it stores. The data thereby obtained by the replacement node is a $[6,4]$-MDS code with the message as $\{a+b,\ b+c,\ d+e,\ e+f\}$; the six symbols obtained by the replacement node are \\ 
\begin{itemize}
\item $a+b\qquad\qquad\qquad\qquad\qquad\qquad\qquad\left(=(a+b)+0(b+c)+0(d+e)+0(e+f)\right)$
\item $a + 3b + 9d + e	 + b + 3c + 9e + f~\qquad\left(=(a+b)+3(b+c)+9(d+e)+(e+f)\right)$
\item $a + 2b + 4d + 8e	+  b + 2c + 4e + 8f~\quad\left(=(a+b)+2(b+c)+4(d+e)+8(e+f)\right)$
\item $a + 6b + 10d + 8e + b + 6c + 10e + 8f~\left(=(a+b)+6(b+c)+10(d+e)+8(e+f)\right)$
\item $a + 5b + 12d + 8e  + b + 5c + 12e + 8f~\left(=(a+b)+5(b+c)+12(d+e)+8(e+f)\right)$
\item $a + 4b + 3d + 12e + b + 4c + 3e + 12f~\left(=(a+b)+4(b+c)+3(d+e)+12(e+f)\right)$
\end{itemize}
The $[6,4]$ MDS property allows for correction of one arbitrary error. It follows that the replacement node can correctly recover $\{a+b,\ b+c,\ d+e,\ e+f\}$ and compute its desired data $(a+b+d+e,\ b+c+e+f)$ even in the presence of an adversary who may gain control of one arbitrary node.
\end{example}

\subsection{Algorithm for Secure Data Reconstruction}
The following theorem presents an explicit algorithm for data-reconstruction that ensures security in the presence of nodes compromised to an active adversary.

\begin{thm}[Secure data-reconstruction]
In the code presented, a data-collector can recover all the $B$ message symbols by downloading data stored in $(k+2p)$ arbitrary nodes, in the presence of up to $p$ compromised nodes. The choice of parameter $p$ is arbitrary, and can be different for every instance of data-reconstruction, thus providing on-demand security.
\end{thm}
\begin{IEEEproof}
Consider a data-reconstruction operation that requires security from up to $p$ compromised nodes. Then, under our protocol for data-reconstruction, a data-collector connects to $(k+2p)$ arbitrary nodes in the system and downloads the $\alpha$ symbols stored in each of these nodes. Let $\bigsetI$ denote this set of $(k+2p)$ nodes. Of these, let $\mathcal{A}$ denote the subset of nodes compromised to the adversary. To prove the correctness of our reconstruction algorithm for the desired protection-level $p$, we assume that the size of $\mathcal{A}$ is no larger than $p$. Of course, the set $\mathcal{A}$ is not known to the decoding algorithm, and is used here only to illustrate the algorithm.
 
\begin{figure}[t]
\centering
\includegraphics[width=.9\textwidth]{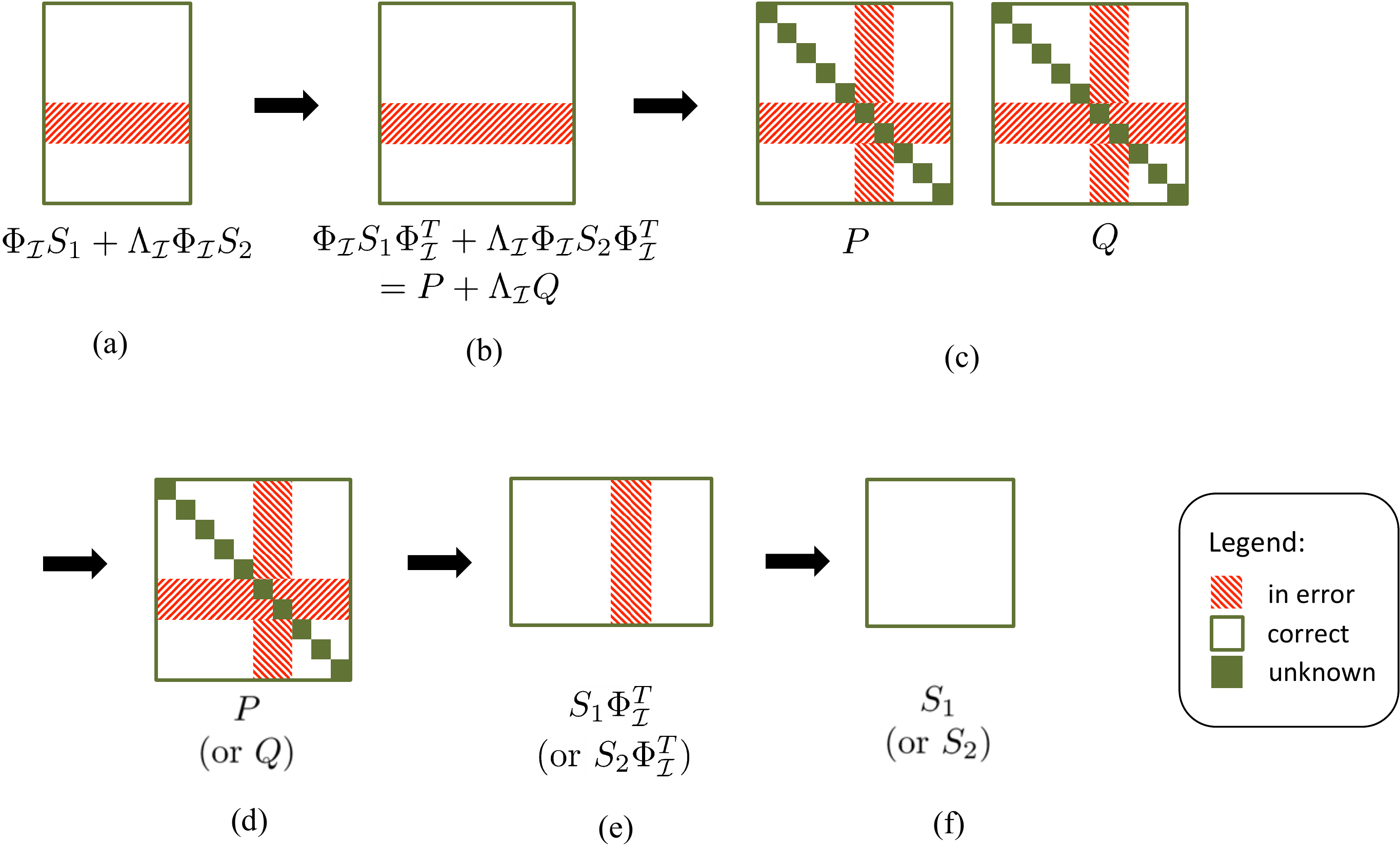}
\caption{An illustration of the error patterns arising during the reconstruction process under the product-matrix MSR code providing security from active adversaries. The pattern shown in part (a) depicts the corruption of some $p$ rows (corresponding to the set $\mathcal{A}$) of the received data. The remaining parts (b)-(f) illustrate the resulting error patterns at various steps of the decoding algorithm. Note that the $p$ corrupt rows are shown as contiguous only for ease of illustration.}
\label{fig:MSR_error_patterns}
\end{figure}

Let $\Psi_\setI$ denote the $((k + 2p) \times d)$ submatrix of $\Psi$, whose rows comprise the set $\mathcal{I}$ of the rows of $\Psi$. From the structure of the encoding matrix $\Psi$~\eqref{eq:MSR_psi}, we can write $\Psi_\setI$ as
\beq \Psi_\setI = \begin{bmatrix}
    \Phi_\setI & \Lambda_\setI\Phi_\setI
    \end{bmatrix}. \nonumber\eeq
\noindent
Then the data collector has access to the encoded symbols  
\bea \Psi_\setI M &=& \begin{bmatrix}
    \Phi_\setI & \Lambda_\setI\Phi_\setI
    \end{bmatrix}
    \begin{bmatrix}
    S_{1} \nonumber \\
    S_{2}
    \end{bmatrix} \\
&=& \begin{bmatrix}
    \Phi_\setI S_{1} + \Lambda_\setI\Phi_\setI S_{2}
   \end{bmatrix}, \eea
of which the $p$ rows corresponding to $\mathcal{A}$ can be corrupt. An example of such an error pattern is illustrated in Fig.~\ref{fig:MSR_error_patterns}a. 

The data-collector multiplies this term by $\Phi_\setI^T$ to obtain
\bea
    \Psi_\setI M \phi_\setI^T &= &\begin{bmatrix}
    \Phi_\setI S_{1} + \Lambda_\setI\Phi_\setI S_{2}
   \end{bmatrix}
    \begin{matrix}
    \Phi_\setI^T
    \end{matrix} \nonumber \\
 &=& \Phi_\setI S_{1} \Phi_\setI^T+ \Lambda_\setI\Phi_\setI S_{2} \Phi_\setI^T~.
\eea
Note that since this operation involves only column operations, the error patterns do not change, and only the $p$ rows corresponding to $\mathcal{A}$ continue to be corrupt. Next, define two $((\alpha+1)\times(\alpha+1))$ matrices $P$ and $Q$ as
\bea
P = \Phi_\setI S_{1} \Phi_\setI^T~, \\
Q = \Phi_\setI S_{2} \Phi_\setI^T~.
\eea
Since $S_{1}$ and $S_{2}$ are symmetric, it follows that the matrices $P$ and $Q$ are also symmetric. One can rewrite the data available to the data-collector in terms of $P$ and $Q$ as
\[ P + \Lambda_\setI Q~,\]
of which the $p$ rows corresponding to $\mathcal{A}$ can be corrupt (Fig.~\ref{fig:MSR_error_patterns}b). 
Now, for any $i$ and $j$ such that $1 \leq i,j \leq k+2p$ and $i\neq j$, the $(i,\ j)^{\text{th}}$, element of the matrix $P + \Lambda_\setI Q$ is
\bea P_{ij} + \lambda_{i}Q_{ij}  , \label{eq:MSR_recon_ij_ele} \eea
and its $(j,\ i)^{\text{th}}$ element is
\bea
&P_{j i} + \lambda_{j}Q_{j i} \nonumber \\
&~~~~~= P_{i j}+\lambda{j}Q_{i j}~, \label{eq:MSR_recon_ji_ele}
\eea
where \eqref{eq:MSR_recon_ji_ele} follows from the symmetry of matrices $P$ and $Q$. By construction, $\lambda_i \neq \lambda_j$ whenever $i\neq j$, and hence using \eqref{eq:MSR_recon_ij_ele} and~\eqref{eq:MSR_recon_ji_ele}, the data-collector solves for the values of $P_{ij}$ and $Q_{ij}$ for all $i \neq j$ (in other words, the data-collector solves for all non-diagonal elements of $P$ and $Q$). However, some of these computed values may be in error due to the corruption of $p$ of the rows in $P + \Lambda_\setI Q$. In particular, the $p$ rows (and hence the $p$ columns) corresponding to $\mathcal{A}$ may be in error in $P$ as well as in $Q$. These error patterns in $P$ and $Q$ are illustrated in Fig.~\ref{fig:MSR_error_patterns}c.

We will now use the (possibly corrupt) values of the non-diagonal elements of matrices $P$ and $Q$ to correctly decode $S_1$ and $S_2$ respectively. Let us first consider recovery of $S_1$, for which we will use the values of the non-diagonal elements of $P$ obtained above. Consider any column $j~(1 \leq j \leq k+2p)$ of the matrix $P$, and let $\mathcal{J}=\mathcal{I} \backslash \lbrace j \rbrace$. The $j^{\text{th}}$ column of the matrix $P$, excluding the element on its diagonal, is 
\[ \Phi_\setJ S_1 {\phi}_j~, \]
 and is of length $(k+2p-1) = (\alpha+2p)$. Recall that by construction, $\Phi$ is the generator matrix of an $[n, \ \alpha]$ MDS code. Hence its submatrix $\Phi_\setJ$ is the generator matrix of an $[\alpha+2p,\ \alpha]$ MDS code. It follows that $\Phi_\setJ$ has a minimum distance of $(2p+1)$. The symbols $\Phi_\setJ S_1 {\phi}_j$ can be considered as an encoding of the vector $S_1 {\phi}_j$ using the generator matrix $\Phi_\setJ$. Using the MDS property of $\Phi_\setJ$, the algorithm attempts to decode $S_1 {\phi}_j$, assuming the presence of no more than $p$ errors. Recall from the discussion above that the vector $S_1 {\phi}_j$ has at most $p$ corrupt entries if $j \notin \mathcal{A}$, but it can be entirely corrupt if $j \in \mathcal{A}$ (see Fig.~\ref{fig:MSR_error_patterns}d).  As a result, the data-collector obtains the correct value of $S_1 \Phi_\setJ$ if $j\notin \mathcal{A}$, and a possibly corrupt value if $j \in \mathcal{A}$ as shown in Fig.~\ref{fig:MSR_error_patterns}e (of course, the data-collector still does not know $\mathcal{A}$ and hence does not know if the value obtained is correct or not). However, when the size of $\mathcal{A}$ is at most $p$, at most $p$ of these decoded vectors $\{ S_1 {\phi}_j \}_{j=1}^{\alpha+2p+1}$ may be in error. Now, these vectors comprise the columns of the matrix $S_1\Phi_\setI^T$ as seen in Fig.~\ref{fig:MSR_error_patterns}e. Again, since $\Phi_\setI$ is the generator matrix of an $[\alpha+2p+1,\ \alpha]$ MDS code, one can decode the $(\alpha \times \alpha)$ matrix $S_1$ from $S_1\Phi_\setI^T$ in the presence of up to $p$ erroneous columns. Thus, the data-collector correctly recovers $S_1$.

Following steps identical to recovery of $S_1$ from $P$, the data-collector also correctly decodes matrix $S_{2}$ from the non-diagonal elements of $Q$. In this manner, the data collector correctly recovers all the message symbols in the presence of at most $p$ compromised nodes.
\end{IEEEproof}

\subsection{Conversion to systematic form}\label{sec:MSR_systematic}
Any MSR code where failed nodes are replaced by their exact replicas can be converted to systematic form through a remapping of the source symbols. An explicit algorithm to convert the product-matrix MSR codes was provided in~\cite{ourProductMatrix}, and is briefly reproduced here for subsequent use. Let $\Psi_k$ denote the $(k \times d)$ submatrix of matrix $\Psi$ with its $k$ rows comprising the first $k$ rows of $\Psi$. Let $U$ be a $(k \times \alpha)$ matrix consisting of the $B\ (=k \alpha)$ message symbols as its $k\alpha$ entries. Now, choose the message matrix $M$ to satisfy
\beq \Psi_k M = U~, \label{eq:secure_MSR_systematic}\eeq
while maintaining the structure of $M$ as given by~\eqref{eq:MSR_M_structure}.

The value of $M$ in~\eqref{eq:secure_MSR_systematic} can be obtained by using the decoding algorithm for data reconstruction (for $p=0$). With this $M$ as the message matrix, under the encoding algorithm of Section~\ref{sec:encoding_A_MSR}, the first $k$ nodes will store the data $\Psi_k M$. This is precisely the matrix of \textit{uncoded} message symbols $U$, thus making these nodes systematic. Note that following this remapping, the node repair algorithm remains the same. Data reconstruction from the $k$ systematic nodes requires no computation, while decoding from parity nodes requires the additional step of recovering $U$ from $M$ by pre-multiplication with $\Psi_k$ as in~\eqref{eq:secure_MSR_systematic}. 

\begin{example}\label{ex:systematic}
Fig.~\ref{fig:adv_msr_sys} depicts an example of a \textit{systematic} product-matrix MSR code providing (on-demand) security from malicious adversaries. This code is obtained by converting the code of Example~\ref{ex:MSR_encoding_nonsystematic} (Fig.~\ref{fig:adv_msr_nonsys}) into a systematic form by remapping the source symbols.
\end{example}

\begin{figure*}
\begin{center}
\medmuskip=0\medmuskip
\renewcommand{\arraystretch}{\codeTableRowspace}
\begin{tabular}{|@{\colortablecell}c|ccccccc|ccccccccccc|}
\hline
Node & \multicolumn{18}{@{\colortablecell}c|}{Data Stored\qquad\qquad\qquad\qquad}\\
\hline
Node 1& a & & & & & & & b & & & & & & & & & & \\
Node 2& c & & & & & & & d & & & & & & & & & & \\
Node 3& e & & & & & & & f & & & & & & & & & & \\
Node 4& 6a &+& 8b &+& 2c &+& 6e & 6a &+& 4b &+& 3c &+& 11d &+& 4e &+& 10f\\
Node 5& 6a &+& 4b &+& 11c &+& 10e & 3a &+& 5b &+& 8c &+& 9d &+& 2e &+& 12f\\
Node 6& 6a &+& 9b &+& c &+& 7e & 10a &+& 7b &+& 5c &+& 3d &+& 11e &+& 5f\\
Node 7& 2a &+& 4b &+& 5c &+& 7e & 3a &+& b &+& 8c &+& 3d &+& 2e &+& 9f\\
\hline
\end{tabular}
\end{center}
\caption{A \textit{systematic} MSR code for $(n=7,\ k=3,\ d=4)$. The message is $\{a,b,c,d,e,f\}$ and the finite field of operation is $\mathbb{F}_{13}$. \advMSR{7}{4}{~3}{6}}
\label{fig:adv_msr_sys}
\end{figure*}

\subsection{Shortening for $d \geq 2k-2$}
\label{sec:MSR_extension}
A shortening technique under the MSR regime for constructing low-redundancy codes from the high-redundancy codes was introduced in~\cite{ourITW,ourMISERJournal}. This technique is employed in~\cite{ourProductMatrix} to extend the $d=2k-2$ product-matrix MSR code to the parameter range of $d \geq 2k-2$. In this section we construct secure MSR codes for $d \geq 2k-2$ that meet~\eqref{eq:MSR_parameters} by applying this shortening technique to the secure MSR codes of Section~\ref{sec:encoding_A_MSR}. 

Since the codes are required to provide on-demand security, the shortening procedure to obtain secure codes for $d \geq 2k-2$ is independent of the desired security levels, and is \textit{identical} to the corresponding procedure in the absence of security~\cite[Section V-C]{ourProductMatrix}. We briefly describe this procedure and show that these codes provide optimal security from active adversaries.

Suppose for some positive integer $i$, one wishes to construct a product-matrix MSR code with parameters $[n,\,k,\,d=2k-2+i]$. Moreover, it is desired to construct an optimal code, i.e., a code whose parameters $(B,\,\alpha,\,\beta)$ satisfy~\eqref{eq:MSR_parameters}. As the first step in this task, we construct a product-matrix MSR code for the parameters $[n'=n+i,\,k'=k+i,\,d'=d+i=2k-2+2i]$. Note that this parameter set satisfies $d'=2k'-2$, and one can employ the encoding procedure described in Section~\ref{sec:encoding_A_MSR} for this construction. Next, this code is converted to a systematic form as described in Section~\ref{sec:MSR_systematic}. Let us denote the resultant systematic code by $\mathcal{C}'$. In order to obtain the desired $[n,\,k,\,d=2k-2+i]$ code $\mathcal{C}$, we observe from~\eqref{eq:MSR_parameters} that the number of message symbols $B'$ in code $\mathcal{C}'$ is $(k+i)(k-1+i)$, and the number of message symbols $B$ in code $\mathcal{C}$ is $k(k-1+i)$. Taking a cue from this expression, to obtain code $\mathcal{C}$ from code $\mathcal{C}'$, we set the $i(k-1+i)$ message symbols stored in the first $i$ (systematic) nodes in code $\mathcal{C}'$ to zero. The data stored in the remaining $n$ nodes, which is an encoding of the remaining $B=k(k-1+i)$ message symbols, constitutes the desired $[n,\,k,\,d=2k-2+i]$ code $\mathcal{C'}$.
 
\begin{example}
Fig.~\ref{fig:adv_msrened} depicts an example of a product-matrix MSR code providing (on-demand) security from malicious adversaries for a parameter set satisfying $d>2k-2$. This code is obtained by shortening the code of Example~\ref{ex:systematic} (Fig.~\ref{fig:adv_msr_sys}) by eliminating message symbols $\{a,b\}$ from the code of Example~\ref{ex:systematic}. The message symbols $\{c,d,e,f\}$ in the code of Example~\ref{ex:systematic} (Fig.~\ref{fig:adv_msr_nonsys}) correspond to $\{a,b,c,d\}$ respectively in Fig.~\ref{fig:adv_msrened}.
\end{example}

\begin{figure*}
\begin{center}
\medmuskip=0\medmuskip
\renewcommand{\arraystretch}{\codeTableRowspace}
\begin{tabular}{|@{\colortablecell}c|ccc|ccccccc|}
\hline
Node & \multicolumn{10}{@{\colortablecell} c|}{Data Stored\qquad\qquad\qquad\qquad}\\
\hline
Node 1 &  a	&& &   b	&&&&&&\\
Node 2 &  c	&& &   d	&&&&&&\\
Node 3 &  2a  &+&  6c &   3a  &+&  11b  &+&  4c  &+&  10d\\   
Node 4 &  11a  &+&  10c	 &   8a  &+&  9b  &+&  2c  &+&  12d\\   
Node 5 &  a  &+&  7c	 &   5a  &+&  3b  &+&  11c  &+&  5d	 \\  
Node 6 &  5a  &+&  7c	 &   8a  &+&  3b  &+&  2c  &+&  9d	 \\
\hline
\end{tabular}
\end{center}
\caption{A {systematic} MSR code for $(n=6,\ k=2,\ d=3)$. The message is $\{a,b,c,d\}$ and the finite field of operation is $\mathbb{F}_{13}$. \advMSR{6}{3}{}{5}}
\label{fig:adv_msrened}
\end{figure*}

The following theorem provides explicit algorithms for secure reconstruction and repair algorithms for this code.
\begin{thm}[Secure data-reconstruction and node-repair]\label{thm:xx}
In the code presented, a data-collector can recover all the $B$ message symbols by downloading data stored in $(k+2p)$ arbitrary nodes, and a replacement node can recover all the data stored in the failed node by downloading $\beta=1$ symbols each from $(d+2p)$ arbitrary nodes, in the presence of up to $p$ compromised nodes.  The choice of parameter $p$ is arbitrary, and can be different for every instance of data-reconstruction and node-repair, thus providing on-demand security.
\end{thm}
\begin{IEEEproof}
To see how reconstruction and repair are performed in code $\mathcal{C}$, we can pretend to operate under code $\mathcal{C}'$, and assume that a user (or replacement node) always connects to the first $i$ nodes in addition to the $k$ (or $d$) nodes that it chooses in $\mathcal{C}$. The data in the first $i$ nodes is known to be all zero, and hence cannot be erroneous. Thus, reconstruction or repair in $\mathcal{C}$ with a security level of $p$ is identical to that in $\mathcal{C}'$ with the same security level $p$. It follows that the reconstruction and repair operations in code $\mathcal{C}$ are identical to those in $\mathcal{C}'$.

Optimality is a result of the parameters of $\mathcal{C}$ satisfying the bound~\eqref{eq:MSR_parameters}.
\end{IEEEproof}

\section{MSR codes for Security from Passive Eavesdroppers}\label{sec:E_MSR}
Recall from~\eqref{eq:MSR_parameters} that in the absence of secrecy requirements, the MSR regime has
\beq d\beta = \alpha + (k -1)\beta~.\label{eq:MSR_parameters_alpha_repeat} \eeq
Unlike the MSR codes for security from active adversaries, the codes for security from passive eavesdroppers do not provide on-demand security. Thus, one needs to fix the desired level of security $\{\ell,\,m\}$ at the time of encoding. 


In this section we present explicit constructions of MSR codes for all parameter values $[n, \ k, \ d\geq 2k-2]$ and all $\{\ell,m\}$ providing information-theoretic security from passive eavesdroppers. The $\{\ell,m\}$-secure MSR codes constructed in this paper achieve 
\bea B^* &=& (k-\ell)(\alpha-m\beta)~\label{eq:msr_secret_achieved}\nonumber\\
&=& (k-\ell)(d-k+1-m)\beta~.\label{eq:msr_secret_achieved_expanded}\eea
Note that~\eqref{eq:msr_secret_achieved_expanded} meets the lower bound~\eqref{eq:msr_rawat_bound} for $m \leq 1$ making our codes optimal for this regime and also establishing the secure capacity of this parameter regime.


For a specified set of parameters $[n,\,k,\,d]$, let $R$ denote the difference in the number of message symbols $B$ that can be stored in a system without security, and the number $B^*$ that can be stored in the $\{\ell,\,m\}$-secure system constructed here. From~\eqref{eq:MSR_parameters} and~\eqref{eq:msr_secret_achieved}, the value of $R$ is given by
\bea R &=& B - B^* \nonumber \\ &=& m(k-\ell)\beta + \ell\alpha~.\label{eq:msr_E_random}\eea
We now describe the construction of the secure product-matrix MSR code, which is performed by replacing precisely $R$ message symbols in the code of~\ref{sec:encoding_A_MSR} by $R$ random symbols. As done in Section~\ref{sec:A_MSR}, we will first consider the case of $d=2k-2$, and subsequently extend it to $d\geq 2k-2$ via a shortening procedure.

Observe from~\eqref{eq:MSR_parameters_alpha_repeat} and~\eqref{eq:msr_secret_achieved_expanded} that the parameters $B^*$ and $\alpha$ are multiples of $\beta$. This allows us to perform striping (explained in Section~\ref{sec:PM}) and construct codes for $\beta=1$ without loss of generality.

\subsection{Encoding for $d=2k-2$}\label{sec:encoding_E_MSR}
An $\{\ell,\ m\}$-secure product-matrix MSR code for the parameters $[n,\,k,\,d]$ is obtained by modifying the $[n,\ k,\ d]$ product-matrix MSR code constructed in Section~\ref{sec:encoding_A_MSR}. Let us denote the product-matrix MSR code of Section~\ref{sec:encoding_A_MSR} by $\mathcal{C}$, and the code with security (which will be constructed below) by $\mathcal{C}^*$. 

The code $\mathcal{C}^*$ also belongs to the product-matrix framework, and is described by an $(n \times \alpha)$ code matrix $C^* = \Psi^* M^*$. The matrices $\Psi^*$ and $M^*$ are obtained by modifying the matrices $\Psi$ and $M$ of Section~\ref{sec:encoding_A_MSR} as follows. Choose $\Psi^*$ such that it satisfies the following property in addition to those required for $\Psi$: when restricted to the first $\ell$ columns, any $\ell$ rows of $\Psi^*$ are linearly independent. The choice of $\Psi^*$ as a Vandermode matrix as described in Section~\ref{sec:encoding_A_MSR} satisfies this additional property. 

\begin{figure}[t]
\centering
\includegraphics[width=.4\textwidth]{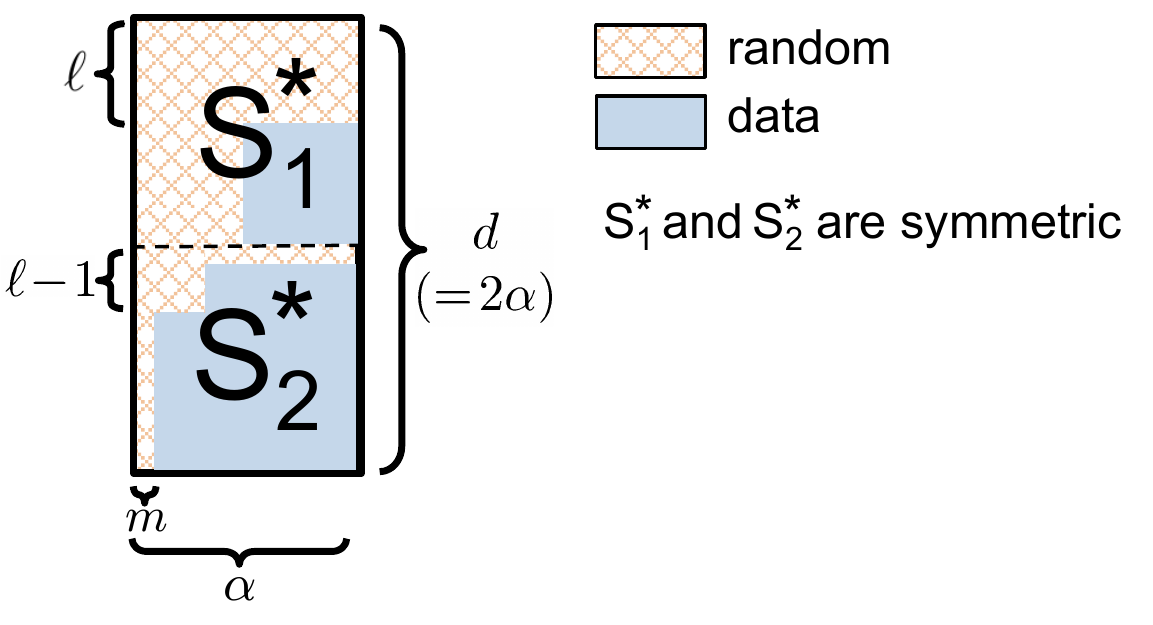}
\caption{Structure of the message matrix $M^*$ for an MSR code secure from passive eavesdroppers, when $d=2k-2$. The matrix $M^*$ for the code providing security from passive eavesdroppers is obtained from the message matrix $M$ of the code without security requirements by filling up a specific set of positions (which were occupied by message values in $M$) by random values. The two submatrices $S_1^*$ and $S_2^*$ are symmetric.}
\label{fig:eaves_M_MSR}
\end{figure}

When $\beta=1$, the value of $R$ in~\eqref{eq:msr_E_random} equals 
$R = \ell \alpha + (k-\ell)m$. The message matrix $M^*$ for code $\mathcal{C}^*$ is obtained by replacing a specific set of $R$ message symbols in $M$ with random symbols. Recall the structure of matrix $M^*$ from~\eqref{eq:MSR_M_structure}. Matrix $M^*$ is obtained from $M$ by replacing: the $\left(\ell\alpha - \frac{\ell(\ell-1)}{2}\right)$ symbols in the first $\ell$ rows (and hence the first $\ell$ columns) of the symmetric matrix $S_1$, the $\frac{\ell(\ell-1)}{2}$ symbols in the \textit{intersection} of the first $(\ell-1)$ rows and first $(\ell-1)$ columns of the symmetric matrix $S_2$, and the $(k - \ell)m$ remaining symbols in the first $m$ rows (and hence the first $m$ columns) of $S_2$. Fig.~\ref{fig:eaves_M_MSR} depicts this procedure.

The following example illustrates this encoding procedure.
\begin{example}
In this example, we will construct an MSR code for $[n=6, \ k=3, \ d=4]$. Let the alphabet of operation be $\mathbb{F}_{13}$. With $\beta=1$,~\eqref{eq:MSR_parameters} leads to having $\alpha=2$ and $B=6$ in the absence of security requirements. Let us suppose we wish to provide security from an eavesdropper who can read the data stored in any one arbitrary node but not read any data downloaded during repair operations. This corresponds to parameters $\ell=1$ and $m=0$. The maximum size of the message that can be stored~\eqref{eq:msr_secret_achieved} is $B^*=4$. We thus choose $B-B^*=2$ symbols $r_1$ and $r_2$ uniformly at random from $\mathbb{F}_{13}$. Let us denote the $4$ message symbols as $\lbrace a, \ b,\ c,\ d \rbrace$. In this code, we choose the matrices $\Psi^*$, $S_1^*$, $S_2^*$ and $M^*$ as
\[
\Psi^* = \begin{bmatrix}
1&0&0&0\\
1 & 1&1&1 \\
1&3&9&1 \\
1&2&4&8\\
1&6&10&8\\
1&5&12&8\\
1&4&3&12
\end{bmatrix}
~~~~,~~
S_1^* = \begin{bmatrix}
r_1 & r_2\\
r_2 & a
\end{bmatrix}
~~~~,~~
S_2^* = \begin{bmatrix}
b & c \\
c & d
\end{bmatrix}
~~~~,~~
M^* = \begin{bmatrix}
S_1\\
S_2
\end{bmatrix}~.
\]
The code is given by $C^*=\Psi^* M^*$. The data stored by the seven storage nodes under this code is depicted in Fig.~\ref{fig:E_MSR_example}.
One can verify that under this code, the data stored in any single node provides no information about the message.
\end{example}

\begin{figure*}
\begin{center}
\medmuskip=0\medmuskip
\renewcommand{\arraystretch}{\codeTableRowspace}
\begin{tabular}{|@{\colortablecell}c|ccccccc|ccccccc|}
\hline
Node & \multicolumn{14}{@{\colortablecell}c|}{Data Stored}\\
\hline
Node 1& $r_1$ & & & & & & &  $r_2$ & & & & & &\\
Node 2&$ r_1$ &+& $r_2$ &+& $b$ &+& $c$	&  $r_2$& +& $a$ & +& $c$& +& $d$\\
Node 3& $r_1$ &+& $3r_2$ &+& $9b$ &+& $c$	&  $r_2$& +& $3a$ &+ &$9c$& +& $d$\\
Node 4& $r_1$ &+& $2r_2$& +& $4b$ &+ &$8c$	&  $r_2$ &+& $2a$ &+& $4c$ &+& $8d$\\
Node 5& $r_1$ &+& $6r_2$ &+& $10b$ &+& $8c$ & $r_2$ &+& $6a$ &+& $10c$ &+& $8d$	\\
Node 6& $r_1$ &+& $5r_2$ &+& $12b$ &+& $8c$ &  $r_2$ &+& $5a$ &+& $12$c &+& $8d$\\
Node 7& $r_1$ &+& $4r_2$ &+& $3b$ &+& $12c$ & $r_2$ &+& $4a$ &+& $3c$ &+& $12d$	\\
\hline
\end{tabular}
\end{center}
\caption{An MSR code for $(n=7,\ k=3,\ d=4)$ providing $\{\ell=1,~m=0\}$ security from passive eavesdroppers: the data is secure from passive eavesdroppers who can read the data \textit{stored} in any one arbitrary node but not to the data passed during repair of any node. The message is $\{a,b,c\}$ and the finite field of operation is $\mathbb{F}_{13}$. The symbols $r_1$ and $r_2$ are drawn uniformly (and independently) at random from $\mathbb{F}_{13}$. The data of any node $f~(1 \leq f \leq 6)$, when required to protect from compromise of any $p=1$ node, is obtained by connecting to any $(d+2p)=6$ nodes and downloading the inner product of the two symbols in each node with $[1~~~x_f]$ where $x_f$ is the $f^{\supth}$ element of $\mathbf{x}=[0~1~3~2~6~5~4]$. The code is optimal in the sense that the amount of storage required is the minimum possible in this setting (it is MDS), and furthermore, the amount of download required for any repair is also the minimum possible under this amount of storage.}
\label{fig:E_MSR_example}
\end{figure*}

\subsection{Reconstruction, Repair and Security}
The following theorems prove the properties of reconstruction, repair and security in the secure PM-MSR code.

\begin{thm}[Data-reconstruction and node-repair]\label{thm:working_E_MSR}
In the code presented, a data-collector can recover all the $B^*$ message symbols by downloading the data stored in any $k$ nodes, and a failed node can be repaired by downloading $\beta=1$ symbol each from any $d$ remaining nodes.
\end{thm}
\begin{IEEEproof}
As in the proof of Theorem~\ref{thm:E_MBR_recon_repair}, treating the random symbols also as message symbols, the secure PM-MSR code $\mathcal{C}^*$ becomes identical to the PM-MSR code $\mathcal{C}$. Thus reconstruction and repair in $\mathcal{C}^*$ are identical to that in $\mathcal{C}$.
\end{IEEEproof}

\begin{thm}[Information-theoretic security]\label{thm:E_MSR_security}
The code $\mathcal{C}^*$ ensures $\{\ell,\ m\}$-security, i.e., an eavesdropper having access to the data stored in up to $\ell$ nodes, and to all the data passed to up to $m$ of these nodes during one or more of their repair operations, gets no information about the message.
\end{thm}
\begin{IEEEproof}
The proof follows the three-step procedure as in the proof in the MBR case (Theorem~\ref{thm:E_MBR_security}). 
Please see the Appendix for the complete proof.
\end{IEEEproof}

\subsection{Shortening for $d\geq 2k-2$}
For any $i\geq 1$, we will now construct product-matrix MSR codes having the parameters $[n,\,k,\,d=2k-2+i]$, for any desired level $\{\ell,\ m\}$ of security. Let us denote this code (which we will construct below) as $\mathcal{C}^*$. The code $\mathcal{C}^*$ will satisfy~\eqref{eq:msr_secret_achieved}, and is optimal when $m\in \{0,1\}$. Throughout, we will assume without loss of generality that the data is striped, and consider $\beta=1$ without loss of generality.

As the first step in this task, a secure product-matrix MSR code is constructed for the parameters $[n'=n+i,\,k'=k+i,\,d'=d+i=2k-2+2i]$, with a security level $\{\ell+i,\ m\}$. Note that this parameter set satisfies $d'=2k'-2$, and one can employ the encoding procedure described in Section~\ref{sec:encoding_E_MSR} for this construction. Denote this code as $\mathcal{C}'^*$.

Our goal is to derive the desired code $\mathcal{C}^*$ by shortening the code $\mathcal{C}'^*$. The shortening procedure described in  Section~\ref{sec:MSR_extension} provides one candidate approach towards this goal. Under this procedure, the code $\mathcal{C}^*$ is derived from $\mathcal{C}'^*$ by setting the data in the first $i$ nodes to zero, and treating all operations in the resultant code $\mathcal{C}^*$ as operations in $\mathcal{C}'^*$. However, this procedure requires the code to be made systematic, which is not possible under a code that provides security, since the compromise of any of the systematic nodes would directly reveal information about the message. Thus, in this section, we will follow a slightly different route to achieve that goal.

Let matrices $\Psi'^*$ and $M'^*$ respectively denote the $(n' \times d')$ encoding matrix and the $(d' \times d')$ message matrix of code $\mathcal{C}'^*$. We will require the matrix $\Psi'^*$ to satisfy an additional property that when restricted to the first $i$ columns, any set of $i$ rows are linearly independent. The choice of $\Psi'^*$ as a Vandermonde matrix, as done in Section~\ref{sec:encoding_A_MSR}, satisfies this condition.

Now, for the specific parameters of codes $\mathcal{C}^*$ and $\mathcal{C}'^*$, observe from~\eqref{eq:msr_secret_achieved} that \beq B^* = B'^* = (k-\ell)( k-1+i-m)~.\nonumber\eeq However, from~\eqref{eq:msr_E_random}, we see that the number of random symbols $R^*$ in code $\mathcal{C}^*$ is \beq R^* = \ell(k-1+i) + (k-\ell)m \nonumber\eeq and the number of random symbols $R'^*$ in $\mathcal{C}'^*$ is \bea R'^* &=&  (\ell+i) (k-1+i) + (k-\ell)m \nonumber \\&=&i(k-1+i)+R^* \nonumber\\ &=&i\alpha + R^*~.\nonumber \eea
Here, $\alpha=(k-1+i)=(d-k+1)=(d'-k'+1)$ is the storage capacity per node in both codes $\mathcal{C}^*$ and $\mathcal{C}'^*$.

To obtain code $\mathcal{C}^*$ from code $\mathcal{C}'^*$, we replace a subset of $i\alpha$ random symbols from $M'^*$ by a set of deterministically chosen symbols in the following manner. Let $\Psi'^*_i$ denote the $(i \times d)$ submatrix of $\Psi'^*$ with its rows comprising the first $i$ rows of $\Psi'^*$. Consider the following $i\alpha$ entries in $M^*$: (a) the $i \alpha - {i \choose 2}$ symbols in the first $i$ rows (and columns) of the symmetric matrix $S_1'^*$, and (b) the ${i \choose 2}$ symbols in the \textit{intersection} of the first $(i-1)$ rows and the first $(i-1)$ columns of $S_2'^*$. In the construction of code $\mathcal{C'}^*$ as per the procedure of Section~\ref{sec:encoding_E_MSR}, each of these entries is populated by a randomly chosen value. With the goal of setting the data in first $i$ nodes as zero, we instead choose these $i\alpha$ elements in a deterministic manner: once all other values of $M'^*$ are fixed, we choose these values to satisfy \[ \Psi'^*_i \left[\begin{array}{c}S_1'^*\\S_2'^*\end{array}\right] =0~.\] 
This is feasible because, by construction, the first $i$ columns of $\Psi_i'^*$ are linearly independent.
Let $M^*$ denote the $(d' \times d')$ matrix resulting from this choice.

The encoding matrix $\Psi^*$ of code $\mathcal{C}^*$ is chosen to be an $(n \times d')$ matrix comprising the last $n$ rows of $\Psi'^*$. Finally, the $[n,\,k,\,d=2k-2+i]$ code $\mathcal{C}^*$ with security $\{\ell,\ m\}$ is given by $C^*=\Psi^* M^*$.

\begin{thm}[Data-reconstruction and node-repair]
In the code presented, a data-collector can recover all the message symbols by downloading the data stored in any $k$ nodes, and a failed node can be repaired by downloading $\beta=1$ symbol each from any $d$ remaining nodes.
\end{thm}
\begin{IEEEproof}
To see how reconstruction and repair are performed in code $\mathcal{C}^*$, we can pretend to operate under code $\mathcal{C}'^*$, and assume that a user (or replacement node) always connects to the first $i$ nodes in addition to the $k$ (or $d$) nodes that it chooses in $\mathcal{C}^*$. This is a valid assumption since the data in the first $i$ nodes of $\mathcal{C}'^*$ is known to be all zero. Thus, reconstruction or repair in $\mathcal{C}^*$ is identical to that in $\mathcal{C}'^*$, and is successful since $\mathcal{C}'^*$ is simply an $[n'=n+i,\,k'=k+i,\,d'=2k'-2=d+i]$ product-matrix MSR code.
\end{IEEEproof}

\begin{thm}[Information-theoretic security]\label{thm:eavesdropper_MSRened}
In the code presented, an eavesdropper having access to the data stored in up to $\ell$ nodes, and to all the data passed to up to $m$ of these nodes during one or more of their repair operations, gets no information about the message.
\end{thm}
\begin{IEEEproof}
Please see the Appendix.
\end{IEEEproof}

\section{Necessary and Sufficient conditions to secure any regenerating code from active adversaries}\label{sec:necessary}
The conversion of the product-matrix codes into codes providing on-demand security, as described in Section~\ref{sec:A_MBR} and Section~\ref{sec:A_MSR}, raises a natural question as to whether \textit{any} regenerating code can provide on-demand security. We answer this question by providing a necessary and sufficient condition for the same. To the best of our knowledge, of all codes known to date, the only codes that satisfy this condition are the product-matrix codes~\cite{ourProductMatrix}.
\begin{thm}\label{thm:necessary}
An $[n,\,k,\,d]$ regenerating code satisfying~\eqref{eq:cut_set_bound_equal} can provide on-demand security from active adversaries and satisfy~\eqref{eq:cut_set_bound_secure_active} with equality if and only if it has a repair mechanism that obeys the following condition: during any instance of repair, the data passed by an existing node to the replacement node does not depend on the identities of the other $(d-1)$ nodes helping in this repair.
\end{thm}
\begin{IEEEproof}
\textit{Necessity}: Consider an $[n,\,k,\,d]$ regenerating code, and assume for now that $(n-d)$ is an odd number. Consider repair of a failed node $f$. Since the code provides on-demand security, it must be able to offer security from the compromise of $p=\frac{n-d-1}{2}$ nodes, by allowing the replacement node to download data from the remaining $(d+2p)=(n-1)$ nodes.  First, observe that all the remaining $(n-1)$ nodes help in the repair, and this removes the dependence on the answer to ``which other nodes help in the repair''. Let $\sigma_1,\ldots,\sigma_{n-1}$ be the data downloaded from the $(n-1)$ remaining nodes. Since arbitrary errors in any $p$ of these must be correctable, it follows that in the absence of any errors, any $d$-sized subset of $\{\sigma_1,\ldots,\sigma_{n-1}\}$ must suffice for correct decoding. Now consider the operation of this code in the absence of security requirements. In this setting, consider the repair scheme wherein a node helping in the repair of a failed node passes precisely what it passed in the case with security. The arguments above ensure that the repair will be successful, and the data $\sigma_i$ passed is independent of the identities of the other $(d-1)$ nodes helping in this repair. If the parameters of the secure code satisfy~\eqref{eq:cut_set_bound_secure_active} with equality, then those of the code without security will satisfy~\eqref{eq:cut_set_bound_equal}. 

Recall the error detection requirement of on-demand security discussed in Remark~\ref{rm:error_detect}. When $(n-d)$ is even, we invoke this requirement of detection of errors from the compromise of any $p=n-d-1$ nodes, wherein the replacement node downloads data from the remaining $(d+p)=(n-1)$ nodes. The rest of the argument is identical to the case of $(n-d)$ being odd considered above.

\textit{Sufficiency:} Consider a regenerating code satisfying the property that in the absence of security requirements, for repair of any node $f$, the data passed by any helper node $h$ is independent of the identities of the remaining $(d-1)$ helper nodes. For a desired security level $p$ during repair of node $f$, let the replacement node connect to some $(d+2p)$ arbitrary nodes. Let each of these $(d+2p)$ nodes pass the data it would have passed to the replacement node in the absence of security requirements. Since the data passed by any $d$ of these nodes would suffice to recover the desired contents in the absence of errors, it follows that the aggregate data from the $(d+2p)$ nodes can correct $p$ arbitrary errors. As a result, this data allows for recovery of the desired data correctly even in the presence of up to $p$ compromise nodes. An identical argument holds for the reconstruction property, thus ensuring that the code provides on-demand security.
If the parameters of the original regenerating code satisfy~\eqref{eq:cut_set_bound_equal} then those of the secure code will satisfy~\eqref{eq:cut_set_bound_secure_active} with equality.
\end{IEEEproof}
\begin{cor}
A code satisfying conditions in Theorem~\ref{thm:necessary} must necessarily operate in the parameter regime $d < (n-1)$.\end{cor}
\begin{IEEEproof}  Clearly, since the number of nodes contacted during repair must satisfy $(d+2p) \leq (n-1)$, the requirement of supporting $p>0$ requires that $n > (d+1)$. It follows that the code should not restrict the number of nodes $n$ to be $(d-1)$.
\end{IEEEproof}
\begin{remark}
The only explicit regenerating codes that support $n>(d+1)$ are the high-rate `approximately-exact' MSR codes of~\cite{ourAllerton} and the product-matrix codes~\cite{ourProductMatrix}. However, the MSR codes of~\cite{ourAllerton} do not satisfy the condition provided in Theorem~\ref{thm:necessary}, while as shown in  Section~\ref{sec:A_MBR} and Section~\ref{sec:A_MSR}, the product-matrix codes do satisfy this condition.
\end{remark}
\section{Discussion}\label{sec:conclusion}
The many recent incidents of the compromise of storage systems (e.g.,~\cite{citibank2011hack,linkedIn2012hack1,yahoo2012hack}) underscore the importance of securing the data in distributed storage systems. Such systems presently employ cryptographic techniques for providing security from passive eavesdroppers. Cryptographic techniques generally assume the eavesdropper to possess only a bounded computational power. However in practice, upon gaining access to the encrypted data, adversaries employ techniques such as intelligent guesswork, dictionary attacks or even crowd-sourcing, and often succeed in decoding significant parts of the data. Information-theoretic security, on the other hand, allows the adversary to posses unbounded computational power, but assumes that the adversary can gain access to only a limited amount of data. In the setting of distributed storage, the information-theoretic security relies on the assumption that the adversary cannot gain access to data stored in more than a certain number of nodes. However, information-theoretic security typically necessitates much greater resource-overheads; in our setting, these are overheads in storage and bandwidth.

The cryptographic and information-theoretic approaches for security rely on different kinds of assumptions, and the security in distributed storage systems may be enhanced by employing both approaches in tandem. To see this, consider a system in which the data is first encoded using conventional cryptographic techniques, following which, an information-theoretically secure erasure code is employed for storing the data and handling node-failures. In such a setting, the adversary first needs to gain access to a certain minimum number of nodes (due to the information-theoretic security), without which it can neither obtain nor corrupt any data; even upon managing to gain such an access, the adversary obtains only the encrypted data. As a result, the layer of information-theoretic security raises the barrier to gain access to the encrypted data. Furthermore, the security provided by the cryptographic encoding allows for the use of only a \textit{thin} layer of information-theoretic security, wherein the threshold on the number of nodes ($\ell$, $m$ or $p$ in our setting) can be as low as $2$ or $3$. This thus aids in overcoming the barrier of the significant overheads faced by information-theoretic security.

An analogy from an end user's perspective is that of storing data (securely) in cloud-based storage services. For instance, one may encode encrypted data using an information-theoretically secure code, and store fractions of it in Dropbox, Google Drive and Microsoft Skydrive storage services.\footnote{dropbox.com, drive.google.com, skydrive.live.com.} Such a code will ensure that a compromise of one of these services provides zero information to the adversary. To get any information about the data, the adversary will need to hack into at least two of the three services. 


In this paper, we addressed the information theoretic part. We presented explicit codes for distributed storage that offer protection from active adversaries and passive eavesdroppers. For a large range of parameters the codes achieve the outer bounds on the system requirements, thus also establishing the capacity of such systems.

\section*{Acknowledgements}
The research of Nihar B. Shah was supported in part by a Berkeley Fellowship and in part by a Microsoft Research PhD fellowship. The research of K. V. Rashmi was supported in part by a Facebook Fellowship and in part by a Microsoft Research PhD Fellowship. This work was also supported in part by NSF grant CIF-0964018 and in part by NSF grant CIF-1116404. The research of P. Vijay Kumar was supported in part by NSF Grant CCF 1421848 and in part by the joint UGC-ISF Research Grant No. 1676/14.

The authors would like to thank Salim El Rouayheb and Sameer Pawar for fruitful discussions.

\appendix
We now present proofs of security from passive eavesdroppers. We use $H(\cdot)$ to denote Shannon entropy and $I(\cdot;\cdot)$ to denote mutual information. All logarithms are taken to base $q$ where $q$ is the size of the finite field of operation of the code under consideration.

\begin{IEEEproof}[\textit{Proof of Theorem~\ref{thm:E_MBR_security}} ($\{\ell,\ m\}$ security under the MBR code)] We consider the worst case where the eavesdropper indeed gains access to the data stored in some $\ell$ nodes and the data passed during any of the repair operations of some $m$ of these $\ell$ nodes. 

Consider the repair of some node $f$ under the product-matrix MBR code, and consider any node $i$ helping in the repair process. Let $\psi_f^*$ and $\psi_i^*$ denote the encoding vectors of nodes $f$ and $i$ respectively. The data passed by node $i$ under the repair algorithm is $(\psi_i^*)^T M^* \psi_f^*$. Since matrix $M^*$ is symmetric, this is equal to $(\psi_f^*)^T M^* \psi_i^*$. Thus, an eavesdropper listening to the data downloaded for repair of node $f$ obtains no more than $(\psi_f^*)^T M^*$, which is the data stored in node $f$.

Let $\Psi_{\text{eve}}^*$ be the $(\ell \times d)$ submatrix of $\Psi^*$, with its $\ell$ rows comprising the $\ell$ rows of $\Psi$ corresponding to the nodes to which the eavesdropper has gained access. Thus the eavesdropper has access to the $\ell d$ symbols in the $(\ell \times d)$ matrix $E^*$ defined as \beq E^* = \Psi_{\text{eve}}^* M^*~.\label{eq:MSR_secret_proof1}\eeq
Let $\mathcal{E}$ denote the set of these $\ell d$ symbols that the eavesdropper has gained access to. Further, let $\mathcal{U}$ denote the set of all $B^*$ message symbols and let $\mathcal{R}$ denote the set of all $R=B-B^*$ random symbols introduced at the encoding stage. In the proof, with some abuse of notation, we will also use the three terms $\mathcal{E}$, $\mathcal{U}$ and $\mathcal{R}$ to denote the random variables corresponding to the respective sets.

As the first step of the proof, we will now show that given the message symbols as side information, an eavesdropper can decode all the random symbols. Under the temporary assumption that the eavesdropper has access to all message symbols, the linearity of the code as in~\eqref{eq:MSR_secret_proof1} allows the eavesdropper to subtract the effect of all message symbols from $E^*$. Denote the resultant $(\ell \times d)$ matrix as $\tilde{E}^*$. Define $\tilde{M}^*$ as a $(d \times d)$ matrix obtained by setting all message symbols in $M^*$ to zero. Thus $\tilde{M}^*$ has its first $\ell$ rows and first $\ell$ columns identical to that of $M^*$, and zeros elsewhere. Then, given the message symbols, the eavesdropper now has access to
 \beq \tilde{E}^* = \Psi_{\text{eve}}^* \tilde{M}^*~.\eeq
Recall the property of $\Psi_{\text{eve}}^*$ that any $\ell$ rows, when restricted to the first $\ell$ columns, are independent. Thus, recovering the $R$ random symbols from $\tilde{E}$ is identical to data reconstruction in a product-matrix MBR code designed for $[\tilde{n} = n, \tilde{k}=\ell, \tilde{d}=d]$ in the absence of security requirements. As a result, given the message symbols, the eavesdropper can decode all the random symbols. Hence we have
\beq H(\mathcal{R}|\mathcal{E},\mathcal{U})=0~.\label{eq:RgivenEU}\eeq

We will now show that $H(\mathcal{E}) \leq R$. From the value of $R$ in~\eqref{eq:mbr_number_random}, it suffices to show that out of the $\ell d$ symbols that the eavesdropper has access to, $\frac{\ell(\ell-1)}{2}$ of them are functions (linear combinations) of the rest. To this end, consider the $(\ell \times \ell)$ matrix
\beq E^* (\Psi^*_{\text{eve}})^T = \Psi^*_{\text{eve}} M^* (\Psi^*_{\text{eve}})^T~. \label{eq:mbr_dependent}\eeq
Since $M^*$ is symmetric, the $(\ell \times \ell)$ matrix in~\eqref{eq:mbr_dependent} is also symmetric. Thus $\ell \choose 2$ dependencies among the elements of $E^*$ can be described by the $\ell \choose 2$ upper-triangular elements of the expression
\beq E^* (\Psi^*_{\text{eve}})^T -\Psi^*_{\text{eve}} (E^*)^T = 0~. \eeq
The linear-independence of the rows of $\Psi^*_{\text{eve}}$ implies that these $\frac{\ell(\ell-1)}{2}$ (redundant) equations are linearly independent. Thus the eavesdropper has access to at most $\left(\ell d - \frac{\ell(\ell-1)}{2}\right)$ independent symbols, i.e., \beq H(\mathcal{E}) \leq R~.\label{eq:EleqR}\eeq

The final part of the proof establishes that given~\eqref{eq:RgivenEU} and~\eqref{eq:EleqR}, the eavesdropper obtains no information about the message.
\begin{eqnarray}
 I(\mathcal{U};\mathcal{E}) &=& H(\mathcal{E}) - H(\mathcal{E} | \mathcal{U}) \label{eq:step3_start} \\
&\leq& R - H(\mathcal{E} |  \mathcal{U})  \label{eq:secrecy_method_1a} \\
&=& R - H(\mathcal{E} |  \mathcal{U}) + H(\mathcal{E} | \mathcal{U},\mathcal{R}) \label{eq:secrecy_method_1b} \\
&=& R - I(\mathcal{E};\mathcal{R} | \mathcal{U})  \\
&=& R - \left( H(\mathcal{R} |  \mathcal{U}) - H(\mathcal{R} |  \mathcal{E},\mathcal{U}) \right) \\
&=& R -  H(\mathcal{R} |  \mathcal{U})  \label{eq:secrecy_method_2} \\
&=& R - H(\mathcal{R})\label{eq:secrecy_method_3}\\
&=& R - R  \label{eq:secrecy_method_4} \\
&=& 0~,\label{eq:step3_stop}\end{eqnarray}
where~\eqref{eq:secrecy_method_1a} follows from~\eqref{eq:EleqR};~\eqref{eq:secrecy_method_1b} is a result of the fact that every symbol in the system is a function of $\mathcal{U}$ and $\mathcal{R}$, resulting in $H(\mathcal{E} | \mathcal{U},\mathcal{R})=0$;~\eqref{eq:secrecy_method_2} follows from~\eqref{eq:RgivenEU}; and~\eqref{eq:secrecy_method_3} follows from the fact that the random symbols are independent of the message symbols.
\end{IEEEproof}

\begin{IEEEproof}[\textit{Proof of Theorem~\ref{thm:E_MSR_security}} ($\{\ell,\ m\}$ security under the MSR code when $d=2k-2$)] We consider the worst case where the eavesdropper indeed gains access to the data stored in some $\ell$ nodes and the data passed during any of the repair operations of some $m$ of these $\ell$ nodes. 
Let $\Psi_{\ell}^*$ be the $(\ell \times d)$ submatrix of $\Psi^*$, with its $\ell$ rows comprising the $\ell$ rows of $\Psi$ corresponding to the nodes to which the eavesdropper has gained access. Due to the specific structure of $\Psi^*$ as $\Psi^* = [\Phi^* ~~~ \Lambda^* \Phi^* ]$, one can write $\Psi_{\ell}^*$ as $\Psi_{\ell}^* = [\Phi_{\ell}^* ~~~ \Lambda_{\ell}^* \Phi_{\ell}^* ]$ where $\Phi_{\ell}^*$ and $\Lambda_{\ell}^*$ are corresponding submatrices of $\Phi^*$ and $\Lambda^*$, and are sized $(\ell \times \alpha)$ and $(\ell \times \ell)$ respectively. Further, let $\Phi_m^*$ be the $(m \times \alpha)$ submatrix of $\Phi^*$, corresponding to the $m$ nodes whose repair operations are also eavesdropped upon. These $m$ nodes are a subset of the set of $\ell$ nodes that constitute the matrix $\Psi^*_{\text{eve}}$, and hence, $\Phi_m^*$ is a submatrix of $\Phi_{\ell}^*$. 

Let $\mathcal{E}$ denote the set of symbols that the eavesdropper has gained access to. Further, let $\mathcal{U}$ denote the set of all $B^*$ message symbols and let $\mathcal{R}$ denote the set of all $R=B-B^*$ random symbols introduced at the encoding stage. In the proof, with some abuse of notation, we will also use the three terms $\mathcal{E}$, $\mathcal{U}$ and $\mathcal{R}$ to denote the random variables corresponding to the respective sets.

Under the repair algorithm of the product-matrix MSR code (see Theorem~\ref{thm:working_E_MSR}), $\mathcal{E}$ comprises the elements of the $(\ell \times \alpha)$ matrix $\Psi^*_{\ell} M^*$ and the elements of the $(d \times m)$ matrix $M^* (\Phi^*_m)^T$.  

Following the approach described in Section~\ref{sec:E_model}, we will first show that given the message symbols as side information, an eavesdropper can decode all the random symbols. Next, using the properties of the matrix $\Psi^*$ and the specific structure of the message matrix $M^*$, we will show that $H(\mathcal{E}) \leq R$. At this point, the arguments in~\eqref{eq:step3_start} to~\eqref{eq:step3_stop} establish that the eavesdropper obtains no information about the message.

\begin{figure}
\centering
\!\!\includegraphics[width=.75\textwidth]{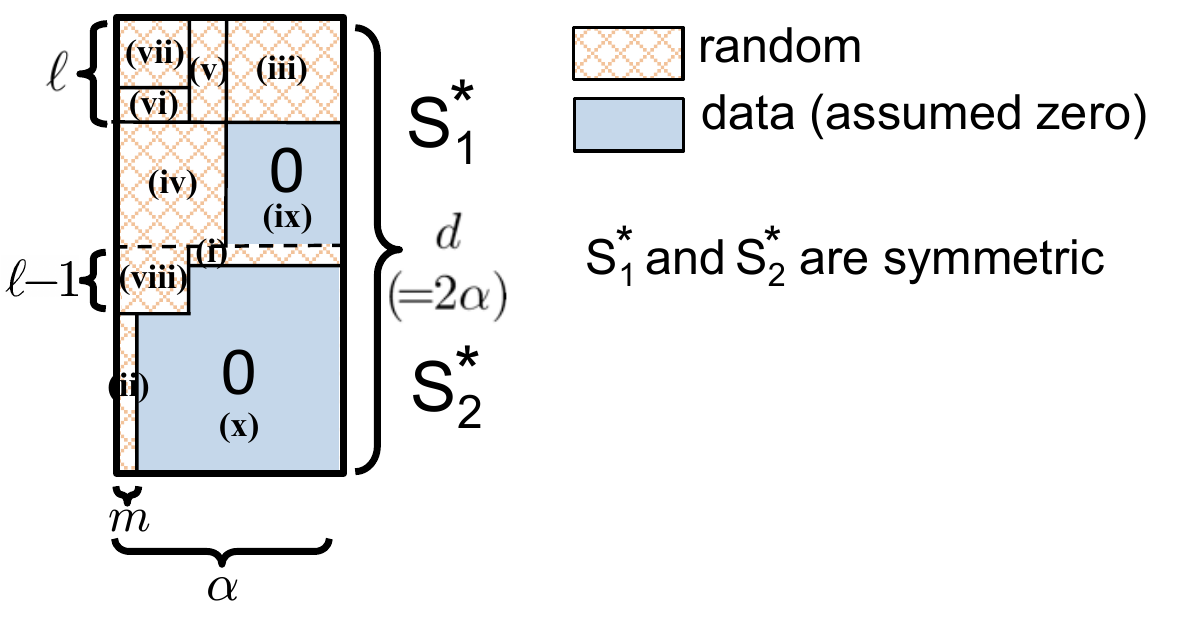}
\caption{Illustration of the order of decoding of the different parts of the message matrix $M^*$ in Step 1 of the proof of security for MSR codes. Note that the parameters considered here satisfy $d=2k-2=2\alpha$.}
\label{fig:E_MSR_security}
\end{figure}

Step 1: In this step, we show that $H(\mathcal{R} | \mathcal{E},\,\mathcal{U})=0$. To this end, assume that a genie provides the values of all message symbols $\mathcal{U}$ to the eavesdropper. Since the code is linear, the eavesdropper can subtract the effects of the message symbols from each of the symbols in $\mathcal{E}$. This allows us to assume (for this step) the values of all message symbols to be zero. Define matrix 
\[ \tilde{M}^* = \left[ \begin{array}{c} \tilde{S_1^*}\\ \tilde{S_2^*}\end{array}\right]\] 
as being identical to $M^*$, but with all message symbols set to zero. It suffices to show that under a code operating with $\tilde{M}^*$ as its message matrix, the eavesdropper can decode all the $R$ random symbols. The eavesdropper has access to the data $\Psi^*_{\ell} \tilde{M}^*$ and $\tilde{M}^* (\Phi^*_m)^T$, and hence equivalently to:
\[ \tilde{E}_1^* = \Phi^*_{m} \tilde{S}_1^*\,\]
\[\tilde{E}_2^* = \Phi^*_{m} \tilde{S}_2^*\,\]
\[\tilde{E}_3^* = \Phi^*_\ell \tilde{S}_1^* + \Lambda_\ell \Phi^*_\ell \tilde{S}_2^*\,\]
where $\tilde{E}_1^*$ and $\tilde{E}_2^*$ are both sized $(m\times (k-1))$ and $\tilde{E}_3^*$ is sized $(\ell \times (k-1))$. By construction, the intersection of the rightmost $(k-\ell)$ columns and the bottommost $(k-1-m)$ rows of the $((k-1)\times(k-1))$ matrix $\tilde{S}_2^*$ contains only the message symbols (depicted as (x) in Fig.~\ref{fig:E_MSR_security}) which are set to zero under our assumptions. Since the rows of $\Phi^*_{m}$ are linearly independent, one can decode the rightmost $(k-\ell)$ columns of $\tilde{E}_2^*$ to recover the symbols in the intersection of top $m$ rows and bottom $(k-\ell)$ columns of $\tilde{S}_2^*$ (depicted as (i) in Fig.~\ref{fig:E_MSR_security}). The symmetry of matrix $S_2^*$ implies the decoding of the intersection of leftmost $m$ columns and bottommost $(k-\ell)$ rows of $\tilde{S}_2^*$ as well (depicted as (ii) in Fig.~\ref{fig:E_MSR_security}). We subtract these symbols out from the remaining data and update the values of $\left\lbrace\tilde{E}_i^* ,~i=1,2,3\right\rbrace$ accordingly. Now, consider the rightmost $(k-1-\ell)$ columns of $\tilde{E}_3^*$. The rightmost $(k-1-\ell)$ columns of $\tilde{S}_2^*$ are either zero or have been subtracted out, and as a result, the rightmost $(k-1-\ell)$ columns of $\tilde{E}_3^*$ are identical to the $(k-1-\ell)$ columns of $\phi_\ell^* \tilde{S}_1^*$. Since the intersection of the bottom $(k-1-\ell)$ rows and the rightmost $(k-1-\ell)$ columns of $\tilde{S}_1^*$ (depicted as (ix) in Fig.~\ref{fig:E_MSR_security}) comprises only of (zero-valued) message symbols, one can decode the rightmost $(k-1-\ell)$ columns of $\tilde{S}_1^*$ from $\phi_\ell^* S_1^*$ (depicted as (iii) in Fig.~\ref{fig:E_MSR_security}). Since $\tilde{S}_1^*$ is symmetric, this also implies decoding of the bottommost $(k-1-\ell)$ rows of $\tilde{S}_1^*$ (depicted as (iv) in Fig.~\ref{fig:E_MSR_security}). Now assume that the values of these decoded symbols are subtracted out from $\left\lbrace\tilde{E}_i^* ,~i=1,2,3\right\rbrace$. Once again we see that the $\ell\supth$ column of $\tilde{S}_2^*$ comprises of symbols that are either zero-valued or have been subtracted out in the previous steps. As a result, the $\ell\supth$ column of $\tilde{E}_3^*$ are identical to the $\ell\supth$ column of $\phi_\ell^* \tilde{S}_1^*$. Since the bottommost $(k-1-\ell)$ elements of the $\ell\supth$ column of $\tilde{S}_1^*$ have been subtracted out, one can decode the $\ell\supth$ column of $\tilde{S}_1^*$ from the $\ell\supth$ column of $E_3^*$ (depicted as (v) in Fig.~\ref{fig:E_MSR_security}). Since $\tilde{S}_1^*$ is symmetric, this also implies decoding of the $\ell\supth$  row of $\tilde{S}_1^*$ (depicted as (vi) in Fig.~\ref{fig:E_MSR_security}). Now assume that the values of these decoded symbols are also subtracted out from $\left\lbrace\tilde{E}_i^* ,~i=1,2,3\right\rbrace$. We continue to focus on $\tilde{E}_3^*$ and observe that the only non-zero elements remaining are the first $(\ell-1)$ rows and columns. Obtaining the top-left $((\ell-1)\times(\ell-1))$ submatrices of $\tilde{S}_1^*$ and $\tilde{S}_2^*$ (depicted as (vii) and (viii) in Fig.~\ref{fig:E_MSR_security}) from this data is identical to the reconstruction requirement of an MSR product-matrix code with parameters $k=\ell$, $d=2\ell-2$ and $\alpha = \ell-1$ in the absence of secrecy requirements. We already know~\cite{ourProductMatrix} that this is possible, and this completes the proof of this step.

Step 2: We will now show that $H(\mathcal{E}) \leq R$. The eavesdropper obtains
\[ {E}_1^* = \Phi^*_{m} {S}_1^*\,\]
\[{E}_2^* = \Phi^*_{m} {S}_2^*\,\]
\[{E}_3^* = \Phi^*_\ell {S}_1^* + \Lambda_\ell \Phi^*_\ell {S}_2^*\,,\]
where ${E}_1^*$ and ${E}_2^*$ are both sized $(m \times (k-1))$, and ${E}_3^*$ is sized $(\ell \times (k-1))$.
These three matrices together consist of \beq |\mathcal{E}| = 2m  (k-1) + \ell  \alpha \eeq symbols. Recall from~\eqref{eq:msr_E_random} that $R=\ell \alpha + m (k-\ell)$. It thus suffices to show that least \bea |\mathcal{E}| - R  &=& (2m  (k-1) + \ell  \alpha) - (\ell \alpha + m (k-\ell)) \\ &=& m(k+\ell-2) \label{eq:redundancy_required}\eea 
of the symbols in $\left\{{E}_1^*,{E}_2^*,{E}_3^*\right\}$ are redundant. 

To this end, first observe that the symmetry of matrices ${S}_1^*$ and ${S}_2^*$ makes the matrices ${E}_1^* (\Phi^*_{m})^T$ and ${E}_2^* (\Phi^*_{m})^T$ also symmetric. As a result, in each of the matrices ${E}_1^*$ and ${E}_2^*$, $\frac{m(m-1)}{2}$ of the symbols are redundant. Now consider matrix ${E}_3^*$. The fact that $\Phi^*_{m} $ is a sub-matrix of $\Phi^*_{\ell}$ implies that the matrix 
\[
{E}_1^* + \Lambda_{m} {E}_2^* = \Phi^*_{m} {S}_1 + \Lambda_{m} \Phi^*_{m} {S}_2\]
comprises $m$ rows of ${E}_3^*$. Thus the $m \alpha = m (k-1)$ symbols in these $m$ rows of $E_3^*$ are redundant. Further, the relation 
\bea {E}_3^* \left( \Phi^*_{m} \right)^T &=& \Phi^*_\ell \left( {E}_1^*  \right)^T + \Phi^*_\ell \left( {E}_2^*  \right)^T\\ &=& \Phi^*_{\ell} {S}_1  \left( \Phi^*_{m} \right)^T +  \Phi^*_{\ell} {S}_2  \left( \Phi^*_{m} \right)^T \eea
implies that $m$ symbols in the span of every row of $E_3^*$ are redundant. Discounting the $m$ rows of $E_3^*$ that we previously showed to be redundant, we see that an additional $m(\ell-m)$ symbols of $E_3^*$ are redundant. Adding these quantities, we get that a total of 
\[ 2\frac{m(m-1)}{2}+m(k-1)+m(\ell-m) = m(k+\ell-2) \]
symbols in the data $\left\{{E}_1^*,{E}_2^*,{E}_3^*\right\}$ obtained by the eavesdropper are redundant, thus meeting the required condition~\eqref{eq:redundancy_required}.

Step 3: Having shown that $H(\mathcal{R} | \mathcal{E},\,\mathcal{U})=0$ and $H(\mathcal{E}) \leq R$, the remainder of the proof is identical to the Step 3 of the proof of the MBR case (Theorem~\ref{thm:E_MBR_security}).
\end{IEEEproof}

\begin{IEEEproof}[\textit{Proof of Theorem~\ref{thm:eavesdropper_MSRened}} ($\{\ell,\ m\}$-security under the MSR code when $d>2k-2$)]
Consider an eavesdropper who gains access to the data stored in an arbitrary set of $\ell$ nodes, and in addition, to the data downloaded for repair by some $m$ of these $\ell$ nodes. We will now show that under code $\mathcal{C}^*$, this eavesdropper obtains no information pertaining to the message. To see this, imagine the existence of a set of $i$ additional nodes that store only zeros as their data. Then, one can assume without loss of generality that the eavesdropper also has access to the data stored in these $i$ nodes. Now, the resultant system is identical to one employing code $\mathcal{C}'^*$, with the eavesdropper having access to the data stored in a subset of $(\ell+i)$ nodes, and the data downloaded for repair by $m$ of these $(\ell+i)$ nodes. As shown in Theorem~\ref{thm:E_MSR_security}, in this case, the eavesdropper obtains no information about the message in the system employing code $\mathcal{C}'^*$. Finally, since the two codes $\mathcal{C}'^*$ and $\mathcal{C}^*$ operate on the same message, it follows that the eavesdropper gains no information about the message in code $\mathcal{C}^*$.
\end{IEEEproof}

\bibliographystyle{IEEEtran}


\end{document}